\documentclass[11pt]{article}
    \usepackage[margin=2cm]{geometry}
    \usepackage{times,bbm,microtype}
    \usepackage{amsmath,amsthm,amssymb,stmaryrd,enumitem,xspace,comment,booktabs}

    \usepackage{graphicx} 
    
    \usepackage[usenames, dvipsnames]{xcolor}
    \usepackage{algorithm}
    \usepackage[noend]{algpseudocode}
    \usepackage{thmtools,thm-restate, mdframed, xparse}
    \usepackage{url}
    \usepackage[sort]{cite}
    \usepackage[linktocpage=true,
	pagebackref=true, colorlinks, linkcolor=BrickRed, citecolor=blue]{hyperref}
    \usepackage{cleveref}
    \usepackage{tikz} 
    \usepackage{caption}
\newcommand{\xqedhere}[2]{\rlap{\hbox to#1{\hfil\llap{\ensuremath{#2}}}}}

\newcommand{\eps}{\varepsilon}

\newcommand{\EE}{\mathbb{E}}

\newcommand{\NN}{\mathbb{N}}

\newcommand{\bx}{\mathbf{x}}
\newcommand{\by}{\mathbf{y}}
\newcommand{\bz}{\mathbf{z}}

\newcommand{\tO}{\widetilde{O}}
\newcommand{\tOmega}{\widetilde{\Omega}}

\renewcommand{\ge}{\geqslant}
\renewcommand{\le}{\leqslant}
\renewcommand{\geq}{\geqslant}

\newcommand{\etal}{{et al.}\xspace}

\DeclareMathOperator{\polylog}{polylog}

\newcommand{\Dout}{\Delta_\text{out}}

\newcommand{\idx}{\textsc{index}\xspace}
\newcommand{\usetint}{\textsc{uniq-setint}\xspace}
\newcommand{\reach}{\textsc{reach}\xspace}
\newcommand{\topo}{\textsc{topo}\xspace}
\newcommand{\ssr}{\textsc{ssreach}\xspace}
\newcommand{\msr}{\textsc{msreach}\xspace}
\newcommand{\reachdag}{\textsc{reach-dag}\xspace}
\newcommand{\ssrdag}{\textsc{ssreach-dag}\xspace}
\newcommand{\acyc}{\textsc{acyc}\xspace}
\newcommand{\sconn}{\textsc{str-conn}\xspace}
\newcommand{\scc}{\textsc{scc}\xspace}
\newcommand{\scctopo}{\textsc{scc-topo}\xspace}
\newcommand{\ssspuw}{\textsc{sssp-uw}\xspace}
\newcommand{\ssspuwdag}{\textsc{sssp-uw-dag}\xspace}
\newcommand{\ssspw}{\textsc{sssp-w}\xspace}
\newcommand{\ssspwdag}{\textsc{sssp-w-dag}\xspace}
\newcommand{\ssspposw}{\textsc{sssp-posw}\xspace}
\newcommand{\ssspposwdag}{\textsc{sssp-posw-dag}\xspace}

\newcommand{\lfdfs}{\textsc{lf-dfs}\xspace}
\newcommand{\lfdfsdag}{\textsc{lf-dfs-dag}\xspace}

\newcommand{\dfs}{\textsc{dfs}\xspace}
\newcommand{\dfsfin}{\textsc{dfs-fin}\xspace}
\newcommand{\dfsfindag}{\textsc{dfs-fin-dag}\xspace}

\newcommand{\sptreeposw}{\textsc{sptree-posw}\xspace}
\newcommand{\sptreeposwdag}{\textsc{sptree-posw-dag}\xspace}
\newcommand{\sptreew}{\textsc{sptree-w}\xspace}
\newcommand{\sptreewdag}{\textsc{sptree-w-dag}\xspace}
\newcommand{\sptreeuw}{\textsc{sptree-uw}\xspace}
\newcommand{\sptreeuwdag}{\textsc{sptree-uw-dag}\xspace}
\newcommand{\longpath}{\textsc{longpath}\xspace}
\newcommand{\longpathsize}{\textsc{longpathsize}\xspace}

\newtheorem{theorem}{Theorem}[section]
\newtheorem{lemma}[theorem]{Lemma}
\newtheorem{conjecture}[theorem]{Conjecture}
\newtheorem{corollary}[theorem]{Corollary}

\newtheorem{fact}[theorem]{Fact}
\newtheorem{proposition}[theorem]{Proposition}

\newtheorem{remark}[theorem]{Remark}
\theoremstyle{definition}

\mdfdefinestyle{thmstyle}{
  backgroundcolor=gray!10,
  linewidth=0pt,
  roundcorner=4pt,
  innertopmargin=\topsep,
  innerbottommargin=\topsep,
  skipabove=\baselineskip,
  skipbelow=\baselineskip
}

\NewDocumentEnvironment{restatablethm}{m m +b}
{%
    \begin{mdframed}[style=thmstyle]
    \begin{restatable}{theorem}{#1}
    \label{#2}
    #3
    \end{restatable}
    \end{mdframed}
}{}

\algrenewcomment[1]{\hfill \textcolor{blue}{$\triangleright$ #1}}
\makeatletter
\algnewcommand{\LineComment}[1]{\Statex \hskip\ALG@thistlm \textcolor{blue}{$\triangleright$ #1}}
\algnewcommand{\LineCommentI}[1]{\Statex \hskip\ALG@thistlm \hskip\algorithmicindent \textcolor{blue}{$\triangleright$ #1}}

\newcommand{\Gscc}{G_\scc}
\newcommand{\DP}{\mathrm{DP}}
\newcommand{\op}{\mathrm{op}}
\newcommand{\indeg}{\mathrm{indeg}}

\title{Square-Root Pass Semi-Streaming Algorithms for Digraph Problems}
\author{Prantar Ghosh\footnote{(pghosh@tntech.edu) Department of Computer Science, Tennessee Tech University, USA. Funded in part by startup support provided by the Department of Computer Science and the College of Engineering at Tennessee Tech.} 
\and 
Amirhossein Janfaza{\footnote{(amirhossein26@ru.is) Department of Computer Science, Reykjav\'{\i}k University, Iceland.}
}
}
\date{}

\begin{document}

\maketitle

\begin{abstract}
We study directed graph problems in the semi-streaming model where we are restricted to $\widetilde{O}(n):=O(n\cdot \text{polylog}(n))$ space for processing the edge stream of an $n$-node graph and can make one or a few passes to solve the given problem. A longstanding open problem in multipass streaming is determining the number of semi-streaming passes needed to solve directed $s,t$-reachability: there is an exponential gap---$\Omega(\log n/\log \log n)$ vs $n^{1/2+o(1)}$---between the best-known lower and upper bounds.
Several related fundamental digraph problems, e.g., topological sorting, acyclicity testing, and decomposition into strongly connected components (SCCs), are known to have the same semi-streaming pass lower bound. However, no $o(n/\text{polylog}(n))$-pass upper bound was known for any of them. Our main result achieves an $n^{1/2+o(1)}$-pass semi-streaming algorithm for each of these problems. As a corollary, we get algorithms with the same pass and space usage for a broad class of DAG problems that admit a certain form of dynamic program over a topological ordering. This class includes weighted single-source shortest paths, longest paths, max-outdegree+1 coloring, and constructing a lexicographically-first DFS (depth-first search) forest. Our algorithm for the last problem can be seen as a generalization of our results on topological sorting and acyclicity testing: recall that standard textbook algorithms for these problems use DFS as a subroutine.

It is easy to show that all our algorithms are optimal up to an $n^{o(1)}$ factor in passes under the reasonable conjecture that $s,t$-reachability requires $\Omega(\sqrt{n})$ semi-streaming passes (weaker than the conjecture that the current-best $n^{1/2+o(1)}$ bound is tight). However, using the fact that our algorithms can be parameterized by the number of passes taken by topological sorting or single-source reachability algorithms, we prove stronger lower bounds for some of the problems even under weaker conjectures. Along this line, we define complexity classes of digraph problems according to their semi-streaming pass complexity and explore the relationships among these classes. Overall, this presents a clearer picture of the hardness of various graph problems in the semi-streaming model.
\end{abstract}

\newpage
\section{Introduction}

%
The semi-streaming model has emerged as a natural setting for processing massive graphs arriving as a stream of edges. Semi-streaming algorithms are restricted to  $\widetilde{O}(n):=O(n\cdot \text{polylog}(n))$ space for an $n$-node graph, which is sublinear in the (worst-case) number of edges, while permitting small amount of information per vertex. This makes many graph problems tractable\cite{FeigenbaumKMSZ05}. When a single pass is insufficient for a solution, a natural workaround is to make more passes over the stream and optimize this number.

\paragraph{Motivation and Background.} The directed $s,t$-reachability problem asks whether there is a directed path from a node $s$ to another node $t$ in the graph. It is one of the most basic problems in the study of graph algorithms, but surprisingly, its complexity is far from resolved in the streaming model. The seminal work of Guruswami and Onak~\cite{GuruswamiO13} proved that any $p$-pass streaming algorithm for the problem requires $n^{1+\Omega(1/p)}/p^{O(1)}$ space. This means any semi-streaming algorithm must make at least $\Omega\left(\log n/\log\log n\right)$ passes. The lower bound holds even when the input graph is promised to be a DAG (directed acyclic graph). This lower bound extends to related directed graph problems. Chakrabarti, Ghosh, McGregor, and Vorotnikova \cite{ChakrabartiGMV20} showed simple reductions from reachability on DAGs to acyclicity and topological sorting. Baweja, Jia, and Woodruff~\cite{BawejaJW22} subsequently showed similar reductions to strong connectivity and decomposition into strongly connected components (SCCs). A couple of works made progress in improving the reachability lower bound, but only for sublogarithmic passes: Assadi and Raz \cite{AssadiR20} proved that almost-linear $n^{2-o(1)}$ space is required for two passes, while \cite{ChenKPSSY21} extended it to hold for $\sqrt{\log n}$ passes. However, their general space/pass tradeoff lower bound did not improve the state of the art for semi-streaming space: the longstanding $\Omega(\log n/\log\log n)$ pass barrier remains.

The best-known upper bounds are nowhere close. Observe that $O(n)$ passes are trivially sufficient: one can simulate DFS (depth-first search) in these many passes (by adding a new node to the DFS forest per pass), which is the textbook approach for all the aforementioned problems in the classical model. To the best of our knowledge, among these problems, only reachability has a published strongly-sublinear-in-$n$ pass upper bound for semi-streaming space: \cite{AssadiJJST22} gave an $n^{1/2+o(1)}$-pass
semi-streaming algorithm, building on the parallel reachability algorithm of Jambulapati, Liu, and Sidford \cite{JLS19}. Their implementation extends to single-source reachability: one can determine all vertices reachable from a source $s$, and moreover recover an explicit directed $s$--$v$ path for every reachable vertex $v$. 

There has been recent work on topological ordering in the streaming model: Chen, Lin, and Tsai \cite{ChenLT25} developed parameterized streaming algorithms for the problem. They show that a DAG with independence number $\alpha$ can be topologically sorted in $O(1/\varepsilon)$ passes using $O(\alpha n^{1+\varepsilon})$ space, and an approximate topological ordering with maximum displacement $\delta$ can be leveraged to obtain an $O(\sqrt n)$-pass algorithm using $O(n+\delta^2)$ space. Indeed, these algorithms do not admit sublinear space for all DAGs (since $\alpha$ and $\delta$ can be $\Theta(n)$ in the worst case). For arbitrary DAGs, they give a tradeoff of
$n/k$-passes and $O(nk)$ space (previously such an algorithm was given by Janthong \cite{Janthong14}). Their framework also yields parameterized algorithms for SCC decomposition, strong connectivity, and related problems. More recently, Chen, Cheung, Lin, and Tsai \cite{ChenCLT26} studied directed strong-connectivity certificates parameterized by the independence number. In the insertion-only model, for instance, they obtain $p$-pass algorithms using roughly $\widetilde{O}(\alpha n^{1+1/p})$ space for a $1$-strong-connectivity certificate, from which several directed connectivity tasks can be solved. Unfortunately, none of these results give $o(n/\polylog n)$-pass semi-streaming algorithms in the worst case.

\paragraph{Our Contributions with Comparison to Prior Work.} Our first contribution is to match the reachability upper bound for topological sorting and SCC-decomposition. We give $n^{1/2+o(1)}$-pass semi-streaming algorithms for the problems on arbitrary digraphs. As immediate corollaries of our topological-sorting and SCC-decomposition algorithms, we obtain the same bound for cycle detection and cycle finding respectively. As mentioned above, each of these problems is known to be at least as hard as $s,t$-reachability on DAGs. Thus, an improvement over our algorithms would, in particular, break the square-root barrier for reachability, at least for DAGs, and maybe shed light on the analogous square-root-depth barrier in parallel reachability, a more longstanding open problem.

We then establish topological sorting as a powerful tool for designing $n^{1/2+o(1)}$-pass semi-streaming algorithms for problems on DAGs. Many classical algorithms on DAGs can be solved by the following approach: first obtain a topological ordering, then evaluate a dynamic program (DP) along this ordering. In the static setting, the implementation is straightforward once the DP is figured out. In a graph stream, however, even knowing the ordering does not immediately make the dynamic program implementable: a vertex may have too many incident edges to store, while its predecessors and successors are scattered arbitrarily throughout the stream. We show that a broad class of such dynamic programs can nevertheless be simulated in $O(\sqrt n)$ semi-streaming passes, once a topological ordering is available. Combining this framework with our topological-sorting algorithm gives $n^{1/2+o(1)}$-pass semi-streaming algorithms for several classical DAG problems, almost all of which generalize $s,t$-reachability. These include single-source shortest paths and the global longest-path problem.

Shortest paths are among the most ubiquitous primitives in graph algorithms, while longest paths in DAGs have classical applications to precedence-constrained scheduling and critical-path analysis \cite{KelleyW59}. Exact-distance computation is considerably more challenging in streaming than offline computation. Hence, existing work on these problems in the streaming setting have focused on approximation \cite{KSS16,KonradT25, AssadiHS26}. There are, however, known $O(\sqrt{n})$-pass algorithms for exact shortest path on {\em undirected} graphs \cite{ChangFHT20}. On directed graphs, Elkin \cite{Elkin11} gave $n/k$-pass $O(nk)$-space  algorithms for exact shortest paths, which necessitates $\Omega(n/\polylog(n))$ passes for semi-streaming space. To our knowledge, no $o(n/\polylog(n))$-pass algorithm was previously known for directed graphs, even for unweighted DAGs. Our algorithms work for weighted DAGs, even on ones with negative edge weights: the absence of negative cycles (in fact, any cycle) removes the usual difficulty.

Recall that DFS generalizes the fundamental problems of ayclicity, topological sorting, and SCC decomposition: classical textbook algorithms use DFS to solve these problems \cite{CLRS}. It is therefore natural to ask whether our results can be generalized to perform DFS, i.e., to construct a DFS forest of the input graph, in the streaming model. Khan and Mehta \cite{KhanM19} gave an $n/\polylog(n)$-pass semi-streaming algorithm for constructing DFS trees of undirected graphs that can be improved to $h/\polylog(n)$-pass when the DFS tree has height $h$. \cite{ChangFHT20} improved upon their results by designing $\tO(\sqrt{n})$-pass semi-streaming algorithms, again for undirected graphs. No bound better than the trivial $O(n)$ was known for digraphs, even for DAGs. Moreover, an arbitrary DFS forest is not sufficient for all of the applications mentioned above: in particular, the standard DFS algorithm for topological sorting relies on the ``finish times'' in addition. Consider the most standard simulation of DFS: break ties in a given lexicographic order. We call the DFS forest thus obtained as the \emph{lexicographically first DFS forest} (\lfdfs). It lets us recover the entire traversal, including its finishing times. We show that the \lfdfs of a DAG can be constructed in $n^{1/2+o(1)}$ semi-streaming passes. The key structural ingredient is a non-trivial DP on the topological order that needs care to establish and then implement in semi-streaming space. Interestingly, while DFS has been a primitive for topological sort in the classical setting, we show that in the streaming model, the reverse is a natural solution.

Our final algorithmic problem is of a slightly different flavor. It is the $(\Dout+1)$-coloring problem: every DAG can be properly colored with $(\Dout+1)$ colors, where $\Dout$ is the maximum outdegree. One can show that the problem is at least as hard as $(\Delta+1)$-coloring ($\Delta$ is the max-degree of an undirected graph) but at least as easy as $(\kappa+1)$-coloring ($\kappa$ is the degeneracy of the graph). $(\Delta+1)$-coloring is known to be admit a single-pass semi-streaming algorithm \cite{AssadiCK19a} while $(\kappa+1)$-coloring requires $\Omega(n^2)$ space in a single pass\cite{BeraCG19}. Although we are not aware of any concrete multipass lower bounds for the $(\kappa+1)$-coloring problem when edges cannot repeat,\footnote{There is an $\tOmega(n)$-pass lower bound for when edges can repeat in the stream \cite{BeraCG19}.} it is known that its natural primitives like finding degeneracy value or a {\em degeneracy ordering} requires $\Omega(n^{1/3})$ passes \cite{AssadiGLMM24}. However, no $o(n)$-pass upper bound is known for any of these degeneracy-based problems. We design an $n^{1/2+o(1)}$-pass $(\Dout+1)$-coloring algorithm. Although the problem does not admit a ``stream-friendly'' DP as the other ones, we combine the idea from our DP framework with other tools like $\ell_0$-sampling and succeed in designing the algorithm. Complementing our algorithmic results, we show that unlike $(\Delta+1)$-coloring and $\Delta$-coloring \cite{AKM22}, $(\Dout+1)$-coloring does not admit a single-pass semi-streaming solution and actually requires linear (i.e., $\Omega(n^2)$) space. On the other hand, we also prove that similar to $(\kappa+1)$-coloring, $(\Dout+1)$-coloring requires $\tOmega(n)$ passes when edges can repeat in the stream.

Finally, we explore hardness relative to the fundamental problems of reachability (single-pair and single-source) on DAGs and topological sorting. Under the natural conjecture that $s,t$-reachability needs $\Omega(\sqrt{n})$ passes, it is immediate that our algorithms are optimal up to $n^{o(1)}$ passes. But since our algorithms can be parameterized by the number of passes required for topological sorting and that for single-source reachability, we show stronger lower bounds (tight up to constant or logarithmic factors) for certain problems even under weaker conjectures that topological sorting or single-source reachability require $\Omega(\sqrt{n})$ passes. Overall, our collection of results makes a significant advancement in the digraph streaming literature which has seen much slower progress than its undirected counterpart. We also present a clearer picture about the relative hardness of fundamental digraph problems whose complexity remains unresolved.

\subsection{Our Results}

Our main result is the following theorem on topological sorting.

\begin{restatablethm}{topothm}{thm:topo}
Given a $p$-pass semi-streaming algorithm for single-source reachability, there is an $O(p\log n)$-pass semi-streaming algorithm for topologically sorting a DAG. In particular, there is an $n^{1/2+o(1)}$-pass semi-streaming algorithm for the problem.
\end{restatablethm}

An algorithm for acyclicity follows as an immediate corollary.

\begin{restatable}{corollary}{acyccor}\label{cor:acyc}
    Given a $p$-pass semi-streaming algorithm for single-source reachability, there is an $O(p\log n)$-pass semi-streaming algorithm for testing whether a graph has a directed cycle. In particular, there is an $n^{1/2+o(1)}$-pass semi-streaming algorithm for the problem.
\end{restatable}

Our next main theorem is an algorithm for SCC decomposition on general digraphs. 

\begin{restatablethm}{sccthm}{thm:sccmain}
Given a $p$-pass semi-streaming algorithm for single-source reachability, there is an $O(p\log n)$-pass semi-streaming algorithm on general digraphs for SCC decomposition, and in fact, for returning the SCCs in topological order. In particular, there is an $n^{1/2+o(1)}$-pass semi-streaming algorithm for SCC decomposition.
\end{restatablethm}

The above algorithm immediately implies an algorithm for finding a directed cycle (if one exists).

\begin{restatable}{corollary}{cycfindcor}\label{cor:cycfind}
    Given a $p$-pass semi-streaming algorithm for single-source reachability, there is an $O(p\log n)$-pass semi-streaming algorithm for finding a directed cycle in a digraph, if one exists, or correctly report that none exists. In particular, there is an $n^{1/2+o(1)}$-pass semi-streaming algorithm for the problem.
\end{restatable}

Note that each of the above problems is fundamental to graph algorithms and can be useful primitives in future research on graph streaming. But prior to our work, no non-trivial ($o(n)$-pass) semi-streaming algorithm was known for them for fully general graphs.

As a consequence of our topological ordering algorithm, we obtain semi-streaming algorithms with the same number of passes for a variety of problems on DAGs. We mention the most general of them, from which algorithms for their special but important variants follow.

\begin{restatablethm}{dagprobsthm}{thm:dagprobs}
     Given a $p$-pass semi-streaming algorithm for topological sorting, there are $O(p+\sqrt{n})$-pass semi-streaming algorithms for the following problems on DAGs: 
     \begin{itemize}
         \item Weighted Single-Source Shortest Path Tree (negative weights allowed)
         \item Weighted (Global) Longest Path
         \item Lexicographically-First DFS Forest
         \item $\Dout+1$-coloring
     \end{itemize}
     In particular, there are $n^{1/2+o(1)}$-pass semi-streaming algorithms for the above problems.
\end{restatablethm}

Except $\Dout+1$-coloring, all problems considered in this work can be shown to be at least as hard as $s,t$-reachability on DAGs. Hence, our bounds for the problems are optimal up to an $n^{o(1)}$-factor in passes subject to the natural conjecture that reachability on DAGs requires $\Omega(\sqrt{n})$ passes. However, since we can express our algorithms' pass usage by the number of passes required for topological sort and single-source reachability, we can show stronger bounds for a subset of these problems even under weaker conjectures that single-source reachability on DAGs or the topological sorting problem require $\Omega(\sqrt{n})$ passes: under the former, our algorithms are optimal up to $O(\log n)$ factors, and under the latter, they are optimal to constant factors; see \Cref{sec:complexity} for details.  

\subsection{Other Related Work} 
Apart from the works mentioned above, prior work relevant to ours include that of Laura and Santaroni \cite{LauraS11} who gave an algorithm for SCC decomposition on W-streams (that allow streaming outputs) using $O(n\log n)$ working memory and $O(n)$ passes. Ghosh and Kuchlous \cite{GhoshK24}, and a subsequent work by Ghosh, Kuchlous, Mehra, and Mukhopadhyay \cite{GhoshKMM26} studied reachability, acyclicity, topological ordering, and SCC decomposition on the special class of digraphs called tournaments (oriented complete graphs), showing that there exist single-pass semi-streaming algorithms for the problems on such digraphs. They also gave single-pass $\tO(n+k)$-space algorithms for digraphs that are $k$ edges away from being a tournament. Very recently, \cite{AzarmehrBCGLR26} gave a $n^{1/2+o(1)}$-pass $(2+\eps)$-approximate streaming algorithm for distance to monotonicity and argued about its optimality subject to the $s,t$-reachability bound.

\subsection{Technical Overview}

\paragraph{Topological Sorting.} At a high level, our algorithm for topological sorting follows a divide-and-conquer strategy reminiscent of quicksort. Suppose that, for a given vertex, we can determine its ancestor and descendant sets in the streaming model. A sampled vertex can then be used as a pivot to partition the remaining vertices into three classes: its ancestors, its descendants, and the vertices unrelated to it. These classes have a natural relative order for topological ordering. By sampling multiple pivots simultaneously and combining their reachability information, we obtain a finer partition of the graph into subproblems whose relative ordering can be determined consistently.

The main challenge is to balance the rate at which the subproblems shrink against the resources required to process the sampled pivots. On the one hand, sufficiently aggressive sampling is needed to reduce the size of each subproblem quickly, thereby keeping the recursion depth, and hence the number of passes, small. On the other hand, increasing the number of sampled vertices creates more reachability queries and potentially many more subproblems. To execute these queries in parallel within semi-streaming space, the reduction in subproblem size must compensate for the growth in the number of sampled pivots.

To achieve this balance, we adopt the aggressive sampling scheme of~\cite{JLS19}, in which the sampling probability increases exponentially with the recursion depth. This choice ensures that progressively smaller subproblems are paired with progressively larger pivot sets while keeping the total amount of reachability information manageable. We use the resulting ancestor--descendant relations not only to partition the graph recursively, but also to impose a consistent ordering on the generated subproblems. After sufficiently many recursion levels, the remaining subproblems have a very small size that can be completely stored.

\paragraph{SCC Decomposition.} The SCC decomposition algorithm builds on essentially the same recursive partitioning framework. The key additional observation is that the strongly connected component containing a sampled vertex $s$ can be identified directly from its reachability information as $C(s)=A(s)\cap D(s).$ Thus, whenever a vertex is sampled, we can immediately identify and remove its entire SCC. The remaining vertices are partitioned according to their ancestor, descendant, or unrelated relationship with the sampled vertices, using the same aggressive sampling scheme as in the topological sorting algorithm.

The crucial property of this partitioning is that vertices belonging to the same SCC always have identical relationships with every sampled vertex, and therefore can never be separated into different subproblems. Consequently, the recursion preserves every undiscovered SCC while progressively reducing the sizes of the remaining subproblems. After the recursive phase, the residual subproblems are sufficiently small that we can compute reachability from all remaining vertices in parallel and recover their SCCs from the intersections of their ancestor and descendant sets.

\paragraph{Shortest and Longest Paths via Dynamic Programming.} Many problems on DAGs admit natural dynamic programming algorithms that process the vertices according to a topological ordering. A simple example is computing the length of the shortest path from a fixed source $s$ to every other vertex: when processing a vertex $v$, its distance can be obtained from the already-computed distances of its in-neighbors. A direct attempt to simulate such a dynamic program in the streaming model is to repeatedly scan the edge stream and update the DP values. The difficulty is that the value of a vertex cannot be finalized until the values of all of its relevant predecessors have themselves been finalized. Consequently, a straightforward simulation may require a large number of passes over the stream.

Our main idea for reducing the number of passes is to partition the topological ordering into consecutive blocks of size roughly $\sqrt{n}$. Since a block contains only $O(\sqrt{n})$ vertices, all edges whose endpoints lie inside a single block can be stored explicitly using $O(n)$ space. The blocks can then be processed sequentially. For shortest paths, while scanning the stream for a block $B$, we aggregate for every $v\in B$ the best contribution coming from in-neighbors in previously processed blocks. After the scan, the stored internal edges of $B$ allow us to complete the DP computation for all vertices of the block offline. Thus, one pass suffices per block, yielding $O(\sqrt{n})$ passes overall.

This argument is not specific to shortest paths. In Section~4, we formalize a generic class of DAG dynamic programs in which the contributions of in-neighbors can be incrementally combined by an associative operation. For every problem admitting such a formulation together with a semi-streaming-size representation of its DP state, the same block decomposition gives an $O(\sqrt{n})$-pass semi-streaming algorithm.

\paragraph{DFS.} For the DFS, our key observation is that, in a DAG, the lexicographically first DFS tree rooted at a vertex $r$ can be characterized through lexicographically minimum paths: for every vertex $v$ reachable from $r$, the unique $r$-to-$v$ path in the lexicographically first DFS tree is precisely the lexicographically smallest path from $r$ to $v$. This characterization turns the problem into a dynamic program over the topological ordering, where for each vertex we select the in-neighbor that yields the lexicographically smallest path. Finally, the lexicographically first DFS forest problem follows from the rooted-tree version through a standard reduction.

\paragraph{$(\Dout+1)$-coloring.} For the $(\Dout+1)$-coloring problem, we aim to simulate the standard offline greedy algorithm, which processes vertices in reverse topological order and colors a vertex once all of its out-neighbors have been colored. In the semi-streaming model, however, we cannot store the complete out-neighborhood of every vertex. We therefore again partition the topological ordering into blocks of size roughly $\sqrt{n}$ and process them in reverse order, storing all edges whose endpoints lie in the current block. At this point, all out-neighbors outside the current block have already been colored. Thus, it suffices to obtain, for each vertex in the current block, a sufficiently large list of colors that remain available with respect to these already-colored out-neighbors. We construct such lists using $\ell_0$-sampling sketches. In particular, we show that with high probability we can sample at least one more color than the degree of each vertex within its block. Since only as many of these colors as the degree of the vertex within its block can be occupied by out-neighbors inside the block, the stored internal edges then allow us to complete the greedy coloring of the entire block offline. This yields an $O(\sqrt{n})$-pass algorithm while using only semi-streaming space.

\subsection{Organization}

The rest of the paper is organized as follows. In \Cref{sec:prelims}, we go over preliminary facts and definitions that we use throughout the paper. In \Cref{sec:toposcc}, we prove our two main results on topological sorting and SCC decomposition. In \Cref{sec:dags}, we show the applications of our topological sorting result to algorithms for several DAG-based problems. In \Cref{sec:complexity}, we investigate conditional lower bounds for the problems. Finally, in \Cref{sec:lbs}, we present two lower bounds on the $(\Dout+1)$-coloring problem.

\section{Preliminaries}\label{sec:prelims}

 \subsection{Notation, Terminology, and Model}

\paragraph{Notation and terminology.} 

We use $G=(V,E)$ to denote a directed graph with $n:=|V|$ vertices and $m:=|E|$ edges. For a positive integer $k$, we write $[k]:=\{1,\ldots,k\}$. Unless stated otherwise, all graphs are directed, simple, and unweighted. For a vertex $v\in V$, let \[ N^+(v):=\{u\in V:(v,u)\in E\} \qquad\text{and}\qquad N^-(v):=\{u\in V:(u,v)\in E\} \] denote its sets of out-neighbors and in-neighbors, respectively. We write $d^+(v):=|N^+(v)|$ and $d^-(v):=|N^-(v)|$ for its out-degree and in-degree, and let \[ \Delta_{\mathrm{out}}(G):=\max_{v\in V} d^+(v). \] When the graph is clear from context, we simply write $\Delta_{\mathrm{out}}$. For vertices $u,v\in V$, we write $u\leadsto v$ if there exists a directed path from $u$ to $v$ in $G$.

\paragraph{The semi-streaming model.} This model was introduced by Feigenbaum \etal \cite{FeigenbaumKMSZ05} to study graph problems in the streaming setting where the input is built by a sequence of updates, and the goal is to optimize the information stored to solve a given problem. A semi-streaming algorithm is restricted to $\tO(n)$ bits of space to process the stream, allowing it to store a small amount of information about each node. In this work, we focus on the insert-only model where edges can be only inserted. The input graph and the stream order are assumed to be chosen adversarially (worst-case) and are fixed before the algorithm sees it. The algorithm can make multiple passes over the stream. Its main goal is to optimize the number of passes. The standard assumption is that the stream order remains the same in every pass (our algorithms, however, do not use this property). After the final pass, the algorithm post-processes the stored data to generate an output. For randomized algorithms, the answer is required to be correct with probability at least $2/3$. 

\subsection{Problems}\label{sec:probs}

We formally define the problems that we consider in this paper and the shorthand we use for them. To each of them, the input is a directed graph $G=(V,E)$.

\begin{description}
    \item \reach ($s,t$-Reachability): Given nodes $s,t \in V$, is $t$ reachable from $s$?

  \item \topo (Topological Sorting): Given that $G$ is a DAG, output a topological ordering of $V$.

\item \ssr (Single-Source Reachability): Given a node $s \in V$, output the set of all nodes reachable from $s$.

\item \msr (Multi-Source Reachability): Given a set of nodes $U\subseteq V$, output the set of all nodes reachable from at least one vertex in $U$.

\item \acyc (Acyclicity Testing): Is $G$ acyclic?

\item \sconn(Strong Connectivity): Is $G$ strongly connected, i.e., is each node in $G$ reachable from every other node?

\item \scc (Strongly Connected Components): Output the partition of $V$ into vertex subsets $(V_1,\ldots,V_k)$ such that each $V_i$ induces a strongly connected component of $G$, i.e., a maximal strongly connected subgraph of $G$.

\item \scctopo (Topological Ordering of SCCs): Output a topological ordering of the SCC-DAG $\Gscc$ obtained by contracting each SCC of $G$ into a supernode. 

\item \ssspw (Weighted Single-Source Shortest Path): Given a source node $s \in V$ and edge weights $w:E\to\mathbb{R}$,
output the length of the shortest path from $s$ to every node $v\in V$, where the weight of a path containing edges $e_1,\ldots,e_k$ is given by $\sum_{i=1}^k w(e_i)$
 (shortest-path length from $s$ to $v$ is $\infty$ if $v$ is not reachable from $s$). 
 
 \ssspposw denotes the version where all edge weights are positive and \ssspuw denotes the unweighted version where $w(e)=1$ for all $e\in E$. 

\item \sptreew (Weighted Shortest-Path Tree):  Given a source node $s \in V$ and edge weights $w:E\to\mathbb{R}$, output a shortest path tree rooted at $s$. 

\sptreeposw denotes the version where all edge weights are positive and \sptreeuw denotes the unweighted version where $w(e)=1$ for all $e\in E$.

\item \dfs (Rooted DFS forest): Output a forest obtained by a DFS on $G$ from a given root $r$ with arbitrary tie-breaking. \dfsfin is the version where we also want the finish time of each vertex. 

\item \lfdfs (Lexicographically-First DFS Forest): Given a total lexicographic ordering of $V$, output the forest obtained by the DFS on $G$ that breaks ties by the lexicographic order.

\item \longpath (Longest Path): Given edge weights $w:E\to\mathbb{R}$, output a longest path in $G$. \longpathsize denotes the variant where we just want the length (sum of edge weights) of a longest path. 


\end{description}

For the above problems that are defined on general digraphs, we use the suffix \textsc{-dag} to denote the variant of the problem where the input is promised to be a DAG, e.g., \reachdag, \lfdfsdag, etc.

\subsection{Basic Tools and Facts}

We use the following version of Chernoff bound.

\begin{fact}[Chernoff bound \cite{DubhashiP09}]\label{fact:chernoff}
	Let $X_1,\ldots,X_k$ be independent indicator random variables. Let 
	$X := \sum_{i=1}^k X_i$ and $\mu_L \leq \EE[X] \leq \mu_H$. Then, for any $\eps > 
	0$, 
	\[
	\Pr\left[X >  (1+\eps) \cdot \mu_H\right] \leq \exp\left(-\frac{\eps^2 \cdot 
	\mu_H}{3+\eps}\right) \quad 
	\textnormal{and} \quad \Pr\left[X <  (1-\eps) \cdot \mu_L\right] \leq \exp\left(-\frac{\eps^2 
	\cdot 
			\mu_L}{2+\eps}\right).
	\]
\end{fact}

The $\ell_0$-sampling problem asks to sample an element uniformly at random from the support (non-zero entries) of a vector. We use the following standard subroutine for the problem.

\begin{fact}[$\ell_0$-Sampling \cite{JowhariST2011}]\label{fact:l0sampling}
    Suppose that we are given streaming updates of $\pm1$ to the entries of an $n$-length vector $\bz$. For $0<\delta<1$, there is an $O(\log^2n\log (1/\delta))$-space algorithm that after reading the stream, outputs   an element from the support of $\bz$ with probability $1-\delta$ (or reports that the support is empty).   
\end{fact}

We exploit the following subroutine in our algorithms.

\begin{fact}[Streaming Single-Source Reachability \cite{AssadiJJST22}]\label{fact:ssrrootn}
    There is a randomized $O(n^{1/2+o(1)})$-pass semi-streaming algorithm for \ssr. It can be extended to find one explicit path (may not be the shortest) from the source to any node in the graph.
\end{fact}

\section{Algorithms for Topological Sort and SCC Decomposition}\label{sec:toposcc}

\subsection{Topological Sorting}

\topothm*

The algorithm then proceeds in two phases.

\paragraph{Phase 1: Recursive Partitioning.}
Phase~1 is a recursive algorithm of depth at most $r_{\max} = \log_k n,$
where \(k>1\) is a fixed constant.

At recursion level \(r\), consider a subgraph \(G'\). We independently sample each vertex into a set \(S\) with probability
\[
p_r = \min\left\{1,\; \frac{20 k^{r+1}\log n}{n}\right\}.
\]
For each sampled vertex $s \in S$, we compute its ancestors, descendants, and unrelated vertices in $G'$. We then compute a valid topological ordering of the vertices in $S$; denote this ordering by
\[
s_1, s_2, \ldots, s_t.
\]

Each vertex $v \in V(G') \setminus S$ is assigned a label vector of length $t$, where the $i$-th coordinate is one of $\{\mathsf{A}, \mathsf{D}, \mathsf{U}\}$, indicating whether $v$ is an ancestor of, a descendant of, or unrelated to $s_i$.  
Vertices with identical label vectors are grouped together, forming subgraphs
\[
G_1, G_2, \ldots, G_q.
\]
These subgraphs are ordered lexicographically according to their label vectors $(\mathsf{A} < \mathsf{U} < \mathsf{D})$, and the algorithm recurses independently on each $G_i$. After completing $r_{\max}$ recursion levels, Phase~1 terminates.

\paragraph{Phase 2: Sorting the Remaining Subproblems.}
In Phase~2, we compute the topological sorting of each remaining subproblem in parallel.

\begin{algorithm}[H]
\caption{\textsc{Toposort}$(G, r)$}
\label{alg:phase1}
\begin{algorithmic}[1]

\If{$r = r_{\max}$}

    \State compute the topological sorting of G
\EndIf

\State $S \gets \emptyset$
\State $p_r \gets \min\left\{1,\; \dfrac{20k^{r+1}\log n}{n}\right\}$

\For{each $v \in V(G)$}
    \State add $v$ to $S$ with probability $p_r$
\EndFor

\For{each sampled vertex $s \in S$}
    \State compute $A_G(s)$, $D_G(s)$, and $U_G(s)$
\EndFor

\State let $S = \{s_1, s_2, \ldots, s_t\}$ be a valid ordering of nodes in $S$

\State assign labels to vertices in $V(G)\setminus S$
\Statex \hspace{\algorithmicindent} according to their relation to vertices in $S$

\State partition $V(G) \setminus S$ into groups of vertices
\Statex \hspace{\algorithmicindent} that share the same label vector
\State let $G_1, G_2, \ldots, G_t$ be the subgraphs induced by these groups

\State order the subgraphs $G_1, \ldots, G_t$ lexicographically
\Statex \hspace{\algorithmicindent} according to their label vectors (with $\mathsf{A} < \mathsf{U} < \mathsf{D}$)

\For{$i = 1$ to $t$}
    \State \Call{Toposort}{$G_i, r+1$}
\EndFor

\end{algorithmic}
\end{algorithm}

\subsection*{Correctness and Complexity Analysis}

\begin{lemma}\label{lem:ancestor-bound}
Let \(v\) be a vertex that lies in a subgraph $G'$ at recursion level $r$. Then, with probability at least $1 - 1/n^{10}$,
\[
|A_{G'}(v)| < k^{-r} n
\quad\text{and}\quad
|D_{G'}(v)| < k^{-r} n,
\]
where $A_{G'}(v)$ and $D_{G'}(v)$ denote the sets of ancestors and descendants of $v$ in $G'$, respectively.
\end{lemma}

\begin{proof}
We prove the bound for ancestors; the argument for descendants is symmetric.

The proof is by induction on $r$.  
For $r=0$, the claim is trivial since $|A_G(v)| < n$.

Assume the claim holds at recursion level $r$ with probability $1 - \frac{r}{n^{11}}$. Consider a vertex $v$ that appears in subgraph $G'$ at level $r$, and let $H$ be the subgraph containing $v$ at level $r+1$.
By the induction hypothesis, $|A_{G'}(v)| < k^{-r} n$
with probability at least $1 - \frac{r}{n^{11}}$.

If $|A_{G'}(v)| < k^{-(r+1)} n$, then the claim holds trivially. Otherwise, let
\[
t = |A_{G'}(v)| \ge k^{-(r+1)} n,
\]
and list the ancestors as $u_1, u_2, \ldots, u_t$, ordered so that there is no path from $u_j$ to $u_i$ for $j>i$.

If the algorithm samples at least one vertex among the last $t/k$ vertices of this list, then all earlier ancestors are separated from $v$, and we obtain
\[
|A_H(v)| \le \frac{1}{k} |A_{G'}(v)| \le k^{-(r+1)} n.
\]

The probability that none of these vertices is sampled is at most
\[
(1 - p_r)^t
\;\le\;
\left(1 - \frac{20 k^{r+1}\log n}{n}\right)^t
\;\le\;
\exp\!\left(- \frac{20 k^{r+1} t \log n}{n}\right)
\;\le\;
n^{-20}.
\]

Thus, with probability at least $1 - \frac{r+1}{n^{11}}$, the bound holds at level $r+1$. Taking a union bound over all vertices and recursion levels completes the proof.
\end{proof}

\begin{lemma}\label{lem:correctness}
The output ordering produced by the algorithm is a valid topological ordering of the input graph.
\end{lemma}

\begin{proof}
The peeling algorithm used in Phase~2 produces a valid topological ordering within each subproblem.

It suffices to show that at recursion level \(r\), concatenating the correctly ordered subgraphs
\[
H_1, H_2, \ldots, H_t
\]
according to the lexicographic order of their label vectors yields a valid ordering for the parent graph \(G'\).

Suppose for contradiction that there exist vertices \(u \in H_i\) and \(v \in H_j\) with \(j>i\) such that \(v\) can reach \(u\) in \(G'\). Since each \(H_i\) is internally ordered correctly, \(u\) and \(v\) must belong to different subgraphs.

Let \(p\) be the first index where the label vectors of \(H_i\) and \(H_j\) differ. The only possible label pairs at index \(p\) are \((\mathsf{A},\mathsf{D})\), \((\mathsf{A},\mathsf{U})\), or \((\mathsf{U},\mathsf{D})\). None of these is compatible with a path from \(v\) to \(u\), since reachability relations are transitive with respect to ancestors and descendants.

This contradiction shows that no such pair \((u,v)\) exists, and the concatenation is a valid topological ordering.
\end{proof}

\begin{lemma}\label{lem:passes}
The algorithm can be implemented in the semi-streaming model using $O(n^{1/2+o(1)})$ passes and $\tilde{O}(n)$ space.
\end{lemma}

\begin{proof}
At each recursion level, we use \Cref{fact:ssrrootn} to compute \ssr for all sampled nodes in parallel, which can be done in $O(n^{1/2+o(1)})$.

By Chernoff bounds \Cref{fact:chernoff}, the number of sampled vertices at level $r$ is $O(k^{r+1}\log n)$ with high probability.  
Moreover, by Lemma~\ref{lem:ancestor-bound}, each vertex $v$ has $A(v)$ and $D(v)$ of size at most $O(k^{-r} n)$. Summing over all subproblems at level $r$, the total space usage is $\tilde{O}(k n)$. Also, it is not necessary to construct the full label vectors explicitly for every node which may requires more than $\tilde{O(n)}$ space. Instead, for a sampled node $s$ and a vertex $v$, we include $s$ in $v$'s label vector only if $v$ is an ancestor or a descendant of $s$. This information suffices to determine the relationship between each sampled node and every other vertex. Combining this observation with Lemma~\ref{lem:ancestor-bound}, we conclude that the total space required to construct the label vectors at each recursive level is $\tilde{O}(kn)$.

Since the number of recursion levels is $O(\log_k n)$, Phase~1 uses $O(n^{1/2+o(1)})$ passes in total. For Phase~2, by Lemma~\ref{lem:ancestor-bound}, the total size of the remaining graph (summed over all subproblems) is $O(n)$. We can therefore store all edges of the remaining subproblems explicitly and run any offline algorithm to compute their topological sorting. Since no additional passes are required beyond storing the edges, the overall pass complexity remains $O(n^{1/2+o(1)})$.
\end{proof}

An algorithm for acyclicity testing with the same space and pass usage follows due to the reduction shown by \cite{ChakrabartiGMV20}: given an algorithm for topological sorting, we can run it on the input for acyclicity. If it is a DAG, we get a valid topological ordering w.h.p., and if it is not, the algorithm either raises an error (if it can detect that the DAG promise was broken) or it returns an invalid topological ordering (since a graph containing a cycle cannot have a valid topological ordering). If an error is raised, we immediately know that the graph has a cycle. Otherwise, we make another pass and check if the returned ordering induces a back edge, which determines whether a cycle exists. The additional pass takes $\tO(n)$ space and the asymptotic pass bound remains the same.

\acyccor*

\subsection{SCC Decomposition}

We describe a recursive algorithm for computing the strongly connected components (SCCs) of a directed graph in the semi-streaming model.

The algorithm proceeds in rounds (recursion levels). At the $r$-th recursion level, the algorithm is applied independently to each current subproblem $G' = (V',E')$.

\paragraph{High-level idea.}
At each recursion level, we sample a set of vertices and use reachability information to identify SCCs containing the sampled vertices. The remaining vertices are then partitioned according to their relationships with the sampled vertices, which induces a partial ordering between the resulting subproblems. The algorithm then recurses on each subproblem.

\paragraph{Detailed description.}
The algorithm consists of two phases.

\paragraph{Phase 1: Recursive SCC discovery and partitioning.}

Let $G'=(V',E')$ be a subproblem in the $r$-th recursion. The algorithm performs the following steps:

\begin{enumerate}
    \item \textbf{Sampling.}
    Sample a set of vertices $S \subseteq V'$ independently at random.

    \item \textbf{Reachability computation.}
    For each $v \in S$, compute its set of ancestors $A(v)$ and descendants $D(v)$ in $G'$

    \item \textbf{Identifying SCCs of sampled vertices.}
    For each \(v \in S\), compute its strongly connected component
    \[
    C(v) = A(v) \cap D(v).
    \]

    For every vertex $u \in C(v)$, assign $C(v)$ as the strongly connected component of $u$ and remove all vertices of $C(v)$ from the current subproblem.

    \item \textbf{Labeling remaining vertices.}
    For each vertex
    \[
    u \in V' \setminus \bigcup_{v\in S} C(v),
    \]
    Assign a label vector describing whether $u$ is

    \begin{itemize}
        \item an ancestor of $v$,
        \item a descendant of $v$,
        \item unrelated to $v$
    \end{itemize}

    for each $v \in S$.

    \item \textbf{Partitioning.}
    Partition the remaining vertices into groups according to identical label vectors.

    \item \textbf{Recursion.}
    Each group defines a new subproblem. The algorithm recurses independently on each subproblem.
\end{enumerate}
The number of total rounds in phase 1  is $r_{\max} = \log_k n$.

\paragraph{Phase 2: Final SCC computation.}

After Phase~1 terminates, some vertices may remain in the final subproblems. For each remaining vertex $v$, we compute its ancestor set $A(v)$ and descendant set $D(v)$ by running BFS in parallel. The strongly connected component of $v$ is then determined as $C(v) = A(v) \cap D(v).$ This procedure identifies all remaining strongly connected components.

\subsection*{Pseudocode}

\begin{algorithm}[H]
\caption{\textsc{SCC}$(G,r)$}
\label{alg:scc-phase1}
\begin{algorithmic}[1]

\If{$r > \log_k n$}
 \For{each vertex $v$}
        \State compute $A(v)$ and $D(v)$
        \State $C(v) \gets A(v) \cap D(v)$
        \State assign $C(v)$ as the SCC of $v$
    \EndFor\EndIf
\State $S \gets \emptyset$
\State $p_r \gets \min\left\{1,\; \dfrac{20k^{r+1}\log n}{n}\right\}$

\For{each $v \in V(G)$}
    \State add $v$ to $S$ with probability $p_r$
\EndFor

\For{each $v \in S$}

    \State compute $A(v)$ and $D(v)$

    \State $C(v) \gets A(v) \cap D(v)$

    \For{each $u \in C(v)$}
        \State assign $C(v)$ as the SCC of $u$
    \EndFor

    \State $C \gets C \cup C(v)$

\EndFor

\State assign labels to vertices in $V(G)\setminus C$
\Statex \hspace{\algorithmicindent} according to their relation to vertices in $S$

\State partition $V(G)\setminus C$ into subgraphs
\Statex \hspace{\algorithmicindent} based on label vectors

\State obtain subgraphs $G_1, G_2, \ldots, G_t$

\For{$i = 1$ to $t$}
    \State \textsc{SCC}$(G_i, r+1)$
\EndFor

\end{algorithmic}
\end{algorithm}

\begin{lemma}\label{size}
Let \(v\) be a vertex that lies in a subgraph \(G'\) at recursion level \(r\). Then, with probability at least \(1 - 1/n^{10}\),
\[
|A_{G'}(v)| < k^{-r} n
\quad\text{and}\quad
|D_{G'}(v)| < k^{-r} n,
\]
where \(A_{G'}(v)\) and \(D_{G'}(v)\) denote the sets of ancestors and descendants of \(v\) in \(G'\), respectively.
\end{lemma}

\begin{proof}
The proof is completely analogous to that of Lemma~\ref{lem:ancestor-bound}, except for one step. In the proof of Lemma~\ref{lem:ancestor-bound}, we used the assumption that the input graph is a DAG only for the step where we ordered all ancestors $u_1, \ldots, u_t$ of $v$ so that there is no path from $u_j$ to $u_i$ whenever $j > i$. Here, we modify the argument as follows. Let $C_1, \ldots, C_h$ be the strongly connected components of the induced subgraph $G[A(v)]$, ordered according to a topological ordering of the SCC DAG of $G[A(v)]$. We place the vertices within each component arbitrarily, and arrange the components consecutively to obtain an ordering $u_1, \ldots, u_t$ of the vertices in $A(v)$. Now, if we sample a node $s$ from the last $t/k$ vertices of this ordering, say $s \in C_j$, then all vertices in $C_j$ are removed. Moreover, every vertex preceding $s$ in this order has relation $A$ or $U$ to $s$, which differs from the relation of $v$ to $s$.
\end{proof}

\begin{lemma}
Let $u$ and $v$ be two vertices that belong to the same strongly connected component of $G$. Then throughout the execution of the algorithm, the vertices $u$ and $v$ are never separated into different subproblems.
\end{lemma}

\begin{proof}
Let $u$ and $v$ be two vertices in the same strongly connected component.
Then $u$ is reachable from $v$ and $v$ is reachable from $u$ in $G$.

Consider an arbitrary recursion step applied to a subproblem $G'$
containing both $u$ and $v$. For every sampled vertex $s \in S$, the
algorithm computes the sets $A(s)$ and $D(s)$ inside $G'$ and assigns
label vectors to vertices according to their relations with the sampled
vertices.

Since $u$ and $v$ are strongly connected, for every sampled vertex $s$
the following holds:

\begin{itemize}
\item If $u \in A(s)$ then $s \leadsto u \leadsto v$, so $v \in A(s)$.
\item If $u \in D(s)$ then $u \leadsto v \leadsto s$, so $v \in D(s)$.
\item If $u$ is unrelated to $s$, then $v$ is also unrelated to $s$,
since otherwise reachability between $u$ and $v$ would imply a path
between $s$ and $u$.
\end{itemize}

Therefore $u$ and $v$ receive identical label vectors with respect to all
sampled vertices in $S$. Consequently, the algorithm assigns $u$ and $v$
to the same partition when forming the subproblems.

Moreover, if a sampled vertex $s$ belongs to the same strongly connected
component as $u$ and $v$, then
\[
C(s) = A(s) \cap D(s)
\]
contains both $u$ and $v$, and the algorithm removes them together as a
single component.

Thus $u$ and $v$ remain in the same subproblem throughout all recursion
levels and are never separated by the algorithm.
\end{proof}

\begin{corollary}
The algorithm correctly identifies all strongly connected components of $G$.
\end{corollary}

\begin{lemma}
The algorithm can be implemented in the semi-streaming model using $O(n^{1/2+o(1)})$ passes and $\tilde{O}(n)$ space.
\end{lemma}

\begin{proof}
At each recursion level, we use \Cref{fact:ssrrootn} to compute \ssr for all sampled nodes in parallel, which can be done in $O(n^{1/2+o(1)})$.

By Chernoff bounds \Cref{fact:chernoff}, the number of sampled vertices at level $r$ is $O(k^{r+1}\log n)$ with high probability.  
Moreover, by Lemma~\ref{size}, each vertex $v$ has $A(v)$ and $D(v)$ of size at most $O(k^{-r} n)$. Summing over all subproblems at level $r$, the total space usage is $\tilde{O}(k n)$. Also, similar to the algorithm for topological sorting, it is not necessary to construct the full label vectors explicitly for every node. We can do it in a similar way using only $\tilde{O}(k n)$ space. Since the number of recursion levels is $O(\log_k n)$, Phase~1 uses $O(n^{1/2+o(1)})$ passes in total.

By Lemma~\ref{size}, we can run BFS from all remaining nodes in parallel, and it requires only $O(1)$ passes and at most $O(n)$ space, since the number of ancestors and descendant for each node in each subproblem is $O(1)$. Hence the overall pass complexity and space usage for this phase are $O(1)$ and $O(n)$, respectively.
\end{proof}

Thus, we get the following theorem.

\sccthm*

An algorithm for finding a directed cycle (if exists, or report NONE otherwise) using the same asymptotic space and passes follows from the above theorem: we first run the SCC decomposition algorithm on the input graph. Every SCC is singleton iff it is a DAG. Otherwise, there is an SCC with at least two nodes $s$ and $t$, and it must contain a cycle. To detect the cycle, parallelly find an $s-t$ path and a $t-s$ path using $n^{1/2+o(1)}$ additional passes and $\tO(n)$ space (\Cref{fact:ssrrootn}). Let $u$ be the first vertex in the $t-s$ path that is also in the $s-t$ path. Then $t \leadsto u \leadsto t$ is a cycle.

\cycfindcor*
\section{Algorithms for DAGs via Dynamic Programming}\label{sec:dags}

In this section, we apply our topological ordering framework to obtain algorithms for various DAG problems.

\subsection{An Algorithm for a Generic DP}

We prove the following theorem.

\begin{theorem}\label{thm:dpalg}
    Let $\Pi$ be a problem whose input is an $n$-node streaming DAG $G=(V,E)$, for which a topological ordering $\sigma$ is known. Let $\op: S^\star \to S$ be an operation on some set $S$ that satisfies $\op(a_1,\ldots,a_{k+1}) = \op(\op(a_1,\ldots,a_{k}),a_{k+1})$ for all $k\in \NN$ and all $a_i\in S$. Suppose that $\Pi$ can be determined by an $n$-length vector $\DP$, indexed by the vertices in $V$, which has the following properties. 
    \begin{enumerate}
        \item $\forall v\in V$ with $\indeg(v) >0$: $\DP[v] = \op (f(\DP[u], (u,v)))$, $\forall u: (u,v)\in E$. 
        \item There is a sketch of size $\tO(n)$ that can (explicitly or implicitly) maintain the vector $\DP$ with streaming updates to its entries.
    \end{enumerate}
    Then there is an $O(\sqrt{n})$-pass semi-streaming algorithm that solves $\Pi$.
\end{theorem}

\begin{proof}
We implement the offline DP recurrence in the semi-streaming model by partitioning the vertices into $\sqrt{n}$ blocks $B_1, \dots, B_{\sqrt{n}}$, where each block contains at most $\sqrt{n}$ consecutive vertices in the topological ordering. For each block $B_i$, let $E(B_i) = \{(u,v)\in E : u,v\in B_i\}$ denote its internal edges.

For the first block $B_1$, the algorithm stores all edges in $E(B_1)$ in one pass (there are $O(|B_1|^2)=O(n)$ such edges) and then runs the offline DP recurrence to compute $\DP[v]$ for all $v\in B_1$.

For each subsequent block $B_i$ with $i>1$, the algorithm proceeds in a single pass as follows. It stores all edges in $E(B_i)$ in memory. Simultaneously, for every $v\in B_i$, it maintains an auxiliary value $S(v)$, initially set to a neutral element $\bot$ (i.e., the identity for $\op$). Whenever the stream presents an edge $(u,v)$ such that $u\notin B_i$ and $v\in B_i$, the algorithm updates
\[
S(v) \leftarrow \op\bigl(f(\DP[u],(u,v)),\, S(v)\bigr).
\]
After finishing the pass, all edges inside $B_i$ are available in memory. The algorithm then processes the vertices of $B_i$ in topological order. For each $v\in B_i$, it initializes $\DP[v] = S(v)$ and then, for every $u \in N^-(v) \cap B_i$ (in increasing topological order), it updates
\[
\DP[v] \leftarrow \op\bigl(\DP[v],\, f(\DP[u],(u,v))\bigr).
\]
Thus, $\DP[v]$ becomes the aggregate of $S(v)$ together with all internal in-neighbors.

Correctness follows by induction on the block index $i$. For $i=1$, there are no outside in-neighbors, so the offline computation is exact. Assume all blocks $B_j$ with $j<i$ have been correctly computed. Consider a vertex $v\in B_i$. Since $G$ is a DAG and the blocks follow the topological order, every in-neighbor $u$ of $v$ either lies in an earlier block $B_j$ ($j<i$) or appears earlier than $v$ within the same block $B_i$. For all outside in-neighbors, $\DP[u]$ is final by the induction hypothesis, and the streaming aggregation exactly constructs
\[
S(v) = \op\bigl( f(\DP[u],(u,v)) : u \in N^-(v) \setminus B_i \bigr).
\]
For the in-neighbors inside $B_i$, the offline loop processes them in topological order, so $\DP[u]$ is already correctly computed when needed. The associativity of $\op$ guarantees that the sequential left fold over the union of these two sets produces precisely
\[
\DP[v] = \op\bigl( f(\DP[u],(u,v)) : u \in N^-(v) \bigr),
\]
which matches the required recurrence. Hence, every vertex is updated correctly.

For the space complexity, by Property 2 of the theorem, the sketch storing $\DP$ uses $\tO(n)$ space. Additionally, each block has size at most $\sqrt{n}$, so its internal edge set $E(B_i)$ has size at most $O(n)$; thus, storing one block's edges at a time requires only $O(n)$ space. Finally, since exactly one pass is spent per block and there are $\sqrt{n}$ blocks, the total number of passes is $O(\sqrt{n})$. This yields the claimed $O(\sqrt{n})$-pass semi-streaming algorithm.
\end{proof}

\begin{corollary}\label{cor:applications}
Theorem~\ref{thm:dpalg} can be applied to the following problems on directed acyclic graphs, yielding for each an $O(\sqrt{n})$-pass semi-streaming algorithm:
\begin{itemize}
    \item \ssspw (hence, \ssspuw as well),
    \item \sptreew (hence, \sptreeuw as well),
    \item \longpath and \longpathsize.
    \item \ssr and \msr.
\end{itemize}
\end{corollary}

\begin{proof}
For each problem, we verify that it satisfies the conditions of Theorem~\ref{thm:dpalg}.

\paragraph{\ssspw{} and \ssspuw.}
Let $\DP[v]$ denote the length of the shortest path from the source $s$ to vertex $v$. For any vertex $v\neq s$ with $\indeg(v)>0$, the shortest path to $v$ satisfies the recurrence
\[
\DP[v] = \min_{u\in N^-(v)} \bigl( \DP[u] + w(u,v) \bigr),
\]
where $w(u,v)$ is the weight of edge $(u,v)$ (with $w(u,v)=1$ for the unweighted version). This fits the theorem's framework by setting $S = \mathbb{R} \cup \{\infty\}$, $\op = \min$ (which is associative), and $f(\DP[u], (u,v)) = \DP[u] + w(u,v)$. The sketch stores the vector $\DP$ explicitly, requiring $O(n)$ space.

\paragraph{\sptreew{} and \sptreeuw.}
For each vertex $v$, let $\DP[v]$ store both the shortest-path distance from $s$ to $v$ and the predecessor of $v$ on a shortest path. The distance component satisfies the same recurrence as in the SSSP case:
\[
dist[v] = \min_{u\in N^-(v)} \bigl( dist[u] + w(u,v) \bigr).
\]
To compute $dist[v]$, the algorithm only needs the distances $dist[u]$ of the in-neighbors $u$ of $v$; the actual paths themselves are not required for the recurrence. Once the minimum is determined, the predecessor of $v$ is set to the in-neighbor $u$ that achieves this minimum. Thus, the sketch maintains two arrays: one storing the distance $dist[v]$ for every vertex $v$, and another storing the predecessor pointer $pred[v]$. The distance array is used to perform the DP updates, while the predecessor pointers are used to reconstruct the shortest-path tree at the end. Both arrays require $O(n)$ total space, so the sketch satisfies the space requirement of Theorem~\ref{thm:dpalg}. The tree is then reconstructed from the predecessor pointers by following them from each vertex back to the root $s$.

\paragraph{\longpath{} and \longpathsize.}
Let $\DP[v]$ denote the length of the longest path in $G$ that ends at vertex $v$. For any vertex $v$ with $\indeg(v)>0$, we have
\[
\DP[v] = \max_{u\in N^-(v)} \bigl( \DP[u] + 1 \bigr),
\]
and for a source vertex (with $\indeg(v)=0$), we set $\DP[v]=0$. This fits the theorem by taking $S = \mathbb{Z}_{\ge 0} \cup \{-\infty\}$, $\op = \max$ (which is associative), and $f(\DP[u], (u,v)) = \DP[u] + 1$. The sketch stores the vector $\DP$ explicitly, requiring $O(n)$ total space. The length of the longest path in the entire graph is then $\max_{v\in V} \DP[v]$, and the path itself can be reconstructed by storing a predecessor pointer for each vertex.

\paragraph{\ssr.}
For a fixed source $s$, let $\DP[v]$ be a Boolean value indicating whether $v$ is reachable from $s$. For any vertex $v\neq s$, we have
\[
\DP[v] = \bigvee_{u\in N^-(v)} \DP[u],
\]
where $\bigvee$ denotes logical OR, and $\DP[s]=1$. This fits the theorem by taking $S = \{0,1\}$, $\op = \text{OR}$ (which is associative), and $f(\DP[u], (u,v)) = \DP[u]$. The sketch stores the reachability vector $\DP$ explicitly, requiring $O(n)$ space.

\paragraph{\msr.}
For a set of sources $S \subseteq V$, the multi-source reachability problem asks to determine which vertices are reachable from any source in $S$. This can be reduced to the single-source reachability problem by constructing an auxiliary graph $G'$ as follows. Add a new dummy source $s^*$ and directed edges from $s^*$ to every vertex in $S$. Since every original edge of $G$ is preserved, a vertex $v$ is reachable from $s^*$ in $G'$ if and only if it is reachable from some source in $S$ in $G$. The reduction adds only $O(n)$ edges and one vertex, so the asymptotic pass and space complexities are preserved. Thus, the \ssr algorithm applies directly to \msr.

In each case, the DP vector is of length $n$, can be maintained with streaming updates, and satisfies the required recurrence with an associative operation. Therefore, Theorem~\ref{thm:dpalg} applies to all of these problems.
\end{proof}

Not every problem of interest, however, fits directly into the DP framework of Theorem~\ref{thm:dpalg}. In particular, the \lfdfs-tree and \lfdfs-forest problems do not admit the requirements of Theorem~\ref{thm:dpalg}. Nonetheless, in the next subsection we will show that a similar result can be obtained: we first reduce the forest problem to the single-root tree case, and then prove a structural property of the \lfdfs-tree that allows us to compute it using the block-decomposition and the generic DP approach.

In the last subsection of this section, we turn our attention to the $(\Delta_{\text{out}}+1)$-coloring problem. This problem is even more different from the generic DP setting, as even its standard offline algorithm is not a dynamic program in the usual sense. Instead, the offline algorithm follows a greedy strategy that processes vertices in the reverse topological order, assigning to each vertex the smallest available color not used by its out-neighbors. We will show that the same block-partitioning idea can still be applied. By processing the blocks in reverse topological order and appropriately sampling available colors for each node, we obtain an $O(\sqrt{n})$-pass semi-streaming algorithm for $(\Delta_{\text{out}}+1)$-coloring as well.

\subsection{Lexicographically-First DFS Forest}\label{sec:dfs}

In this subsection, we turn our attention to the problem of computing the \lfdfs-forest of a DAG $G=(V,E)$, where the standard DFS procedure is initialized on the smallest unvisited vertex at each step. While our algorithm in this section is formally designed for computing the \lfdfs-tree rooted at a fixed starting vertex, it actually suffices to solve only this restricted version. Indeed, the general forest problem can be reduced to the tree problem by a simple and standard transformation, which we describe below.

Given an arbitrary DAG $G=(V,E)$ for which we wish to obtain the \lfdfs-forest, we construct an auxiliary graph $G' = (V', E')$ as follows. Let 
\[
V' = V \cup \{r\},
\]
where $r$ is a new dummy root node. The edge set of $G'$ consists of all original edges of $G$, together with a directed edge from $r$ to every vertex in $V$:
\[
E' = E \cup \{(r, v) : v \in V\}.
\]
Furthermore, we extend the global vertex ordering by declaring $r$ to be strictly smaller than every vertex in $V$.

The key observation is that running the \lfdfs-tree algorithm on $G'$ starting from the single root $r$ exactly simulates the iterative forest construction on $G$. When the DFS begins at $r$, it first visits this dummy root. Since $r$ has outgoing edges to all vertices of $V$, its adjacency list, sorted lexicographically, contains every vertex of the original graph. The DFS then proceeds to the smallest unvisited vertex in $V$, say $r_1$, and from there it explores the entire reachable region of $G$ according to the DFS rules, backtracking only when no further unvisited outgoing edges remain. This exhaustive exploration from $r_1$ precisely yields the first tree of the desired \lfdfs-forest.

Once this first tree has been fully explored, the DFS returns to $r$. At this point, $r$ still has unvisited neighbors in its adjacency list (namely, all vertices of $V$ that were not reached in the first component). The DFS thus advances to the next smallest unvisited vertex and initiates the second tree of the forest. This process repeats until every vertex of $V$ has been visited. Consequently, the \lfdfs-tree rooted at $r$ in the auxiliary graph $G'$ consists of the dummy root $r$ with, attached as its children, all the individual DFS trees of the forest, arranged in the correct lexicographic order.

Therefore, if we remove $r$ and all incident edges from the resulting \lfdfs-tree of $G'$, the remaining disconnected subtrees form exactly the \lfdfs-forest of the original graph $G$.

 Now, we describe an algorithm for computing a DFS tree rooted at an arbitrary vertex $r$ in a DAG in $\sqrt{n}$-passes, assuming that a topological ordering of the DAG is known.

The high-level idea is to compute, for every vertex $v\in V\setminus\{r\}$ that is reachable from $r$, the lexicographically minimum path from $r$ to $v$. We then show that the collection of these paths forms exactly the \lfdfs-tree rooted at $r$.

To compute these paths efficiently in the semi-streaming model, we partition the vertices according to the known topological ordering into $\sqrt{n}$ consecutive blocks. We then use a dynamic programming procedure to compute the desired paths. In particular, for every vertex $v$ appearing after $r$ in the topological ordering, we compute the lexicographically minimum path from $r$ to $v$.

The remainder of this section is organized as follows. We first prove that computing the lexicographically minimum paths from $r$ to all reachable vertices is equivalent to computing the \lfdfs-tree rooted at $r$. We then describe how to fit this new problem into the generic DP.

Let $G = (V, E)$ be a directed acyclic graph, and let the vertices be totally ordered by a relation $<$. We write $P <_{\mathrm{lex}} Q$ if the vertex sequence of $P$ is lexicographically smaller than that of $Q$.

Fix a root vertex $r \in V$. Perform a depth‑first search (DFS) starting at $r$, where at every vertex $x$ its outgoing neighbors are examined in strictly increasing order according to $<$. Let $T$ be the resulting DFS tree rooted at $r$, and for any vertex $v$ reachable from $r$, denote by $P_T(r,v)$ the unique path from $r$ to $v$ in $T$.

\begin{lemma}\label{lemma:path}
For every vertex $v$ reachable from $r$, the path $P_T(r,v)$ is the lexicographically smallest path among all simple paths from $r$ to $v$ in $G$.
\end{lemma}
\begin{proof}
Let $v$ be an arbitrary reachable vertex, and let
\[
P = (p_0, p_1, \dots, p_k), \qquad p_0 = r,\; p_k = v,
\]
be the unique path from $r$ to $v$ in the DFS tree $T$.

Take any other simple path
\[
Q = (q_0, q_1, \dots, q_\ell), \qquad q_0 = r,\; q_\ell = v.
\]
If $P = Q$ there is nothing to prove. Otherwise, let $i$ be the smallest index such that $p_i \neq q_i.$
Since both paths start at $r$, we have $i > 1$. Let
\[
x := p_{i-1} = q_{i-1}, \qquad a := p_i, \qquad b := q_i.
\]
Thus $a \neq b$, and both $a$ and $b$ are outgoing neighbours of $x$.

At the moment when DFS processes vertex $x$, it inspects its neighbours in increasing order. It chooses the first neighbour that is still unvisited and adds the edge $x \to a$ to the DFS tree. Therefore one of the following two cases must hold:

\begin{enumerate}
\item \textbf{$b$ is unvisited when $x$ is processed.}  
Since DFS explores neighbors in increasing lexicographic order and chose $a$ over $b$, we must have $a<b$. Thus, at the first position where $P$ and $Q$ differ, the vertex in $P$ is lexicographically smaller than the vertex in $Q$.

\item \textbf{$b$ was already visited before $x$ is processed.}  
Because DFS is recursive, a vertex that is visited before $x$ and is not an ancestor of $x$ must have been completely explored (finished) before the exploration of $x$ begins. Since $x$ lies on the tree path $P$ to $v$, $x$ is an ancestor of $v$ in $T$; hence $x$ is discovered before $v$.
On the other hand, from the suffix of $Q$ we have a path from $b$ to $v$. Since $b$ was already visited and finished before $x$, $v$ is forced to be discovered and finished within the subtree of $b$, thus $v$ is discovered before $x$. This gives the contradiction, which is impossible. Hence this case cannot occur either.
\end{enumerate}

Because $Q$ was an arbitrary simple path from $r$ to $v$, the DFS‑tree path $P_T(r,v)$ is lexicographically minimal. This holds for every reachable vertex $v$, completing the proof.
\end{proof}

By the above lemma, computing the \lfdfs-tree rooted at $r$ is equivalent to computing, for every vertex $v$ reachable from $r$, the lexicographically minimum path from $r$ to $v$. We now show that this task fits precisely into the DP framework of Theorem~\ref{thm:dpalg}.

For each vertex $v$, let $\DP[v]$ store the predecessor of $v$ on the lexicographically minimum path from $r$ to $v$ (with $\DP[r]=\bot$). The lexicographically minimum path to $v$ satisfies the recurrence
\[
\DP[v] = \arg\min_{u \in N^-(v)} (P(u) \circ v),
\]
where $P(u)$ denotes the path obtained by following the predecessor pointers from $u$ back to $r$, and the minimum is taken with respect to the lexicographic ordering of paths.

To apply Theorem~\ref{thm:dpalg}, we define the operation $\op$ and the function $f$ as follows. Let $S$ be the set of all possible paths starting at $r$, and let $\op$ take the node $u$ such that $P(u)$ is the lexicographic minimum over the set of candidate paths. This operation is associative, as taking the minimum over a set is independent of the order in which the elements are aggregated. For an edge $(u,v)$, the function $f(\DP[u],(u,v))$ represents the path $P(u) \circ v$, which is implicitly encoded by setting the predecessor of $v$ to $u$.

A subtle but important point is that the sketch does not store the full vertex sequence of $P(u)$ explicitly for every $u$, as that would require $\Omega(n^2)$ space. Instead, the sketch maintains the partial \lfdfs-tree formed by the current predecessor pointers. Crucially, when the algorithm needs to compare two candidate paths $P(u_1)\circ v$ and $P(u_2)\circ v$, it suffices to compare $P(u_1)$ and $P(u_2)$, which can be done by traversing the maintained parent pointers of the tree. Thus, the entire collection of paths is implicitly represented by the tree, and the sketch meets the space requirement of Theorem~\ref{thm:dpalg} (Property~2). The recurrence clearly satisfies the associative aggregation form required by Property~1.

Therefore, applying Theorem~\ref{thm:dpalg} immediately yields an $O(\sqrt{n})$-pass semi-streaming algorithm for computing the lexicographically minimum paths from $r$ to all reachable vertices. Now, we are ready to state the main results of this subsection.

\begin{theorem}\label{thm:lfdstree}
Let $G=(V,E)$ be a directed acyclic graph for which a topological ordering is known, and let $r\in V$ be an arbitrary root vertex. There exists an $O(\sqrt{n})$-pass semi-streaming algorithm that computes the \lfdfs-tree rooted at $r$.
\end{theorem}

\begin{proof}
By Lemma~\ref{lemma:path}, computing the \lfdfs-tree rooted at $r$ is equivalent to computing the lexicographically minimum path from $r$ to every reachable vertex. As argued above, this problem satisfies the conditions of Theorem~\ref{thm:dpalg}: the recurrence for the predecessor pointers uses an associative arg min operation, and the sketch requires only $O(n)$ space to store the predecessor pointers. Theorem~\ref{thm:dpalg} then yields an $O(\sqrt{n})$-pass semi-streaming algorithm.
\end{proof}

\begin{theorem}\label{thm:lfforest}
Let $G=(V,E)$ be a directed acyclic graph for which a topological ordering is known. There exists an $O(\sqrt{n})$-pass semi-streaming algorithm that computes the Lexicographically-First DFS forest of $G$.
\end{theorem}

\begin{proof}
This follows directly from the reduction described at the beginning of this subsection. Construct the auxiliary graph $G'$ by adding a dummy root $r$ as described above. Applying Theorem~\ref{thm:lfdstree} to $G'$ rooted at $r$ yields the LF-DFS tree of $G'$ in $O(\sqrt{n})$-passes and $O(n)$ space. Removing the dummy root $r$ and its incident edges from this tree leaves exactly the LF-DFS forest of the original graph $G$. The reduction adds only $O(n)$ edges and one vertex, so the asymptotic pass and space complexities are preserved.
\end{proof}

\subsection{$(\Dout+1)$-coloring}

\begin{theorem}
    Let $G=(V,E)$ be a directed acyclic graph for which a topological ordering is known, there is a randomized
    semi-streaming algorithm that computes a proper
    $(\Dout+1)$-coloring of $G$ using
    $O(\sqrt{n})$ passes with high probability.
\end{theorem}

\begin{proof}
Partition the vertices according to the given topological ordering into $\sqrt{n}$ consecutive blocks $B_1,B_2,\ldots,B_{\sqrt{n}},$
where each block contains at most $\sqrt{n}$ vertices. For every block
$B_i$, let $H_i=G[B_i]$ denote the subgraph induced by the vertices of $B_i$.

The algorithm processes the blocks in reverse order. At the beginning of the algorithm, all vertices are uncolored. When processing a block $B_i$, all vertices in blocks appearing after $B_i$ in the topological ordering have already been colored. We construct a list $L(v)$ of available colors for every $v\in B_i$ and then color $H_i$ using these lists.

For every vertex $v$, we maintain
$c\log n \sqrt n$ independent $\ell_0$-sampling sketches, where $c$ is a sufficiently large constant. The sketches are used to sample available colors for $v$.
The sketches are processed in groups of size $c\log n$. Importantly, when processing the next group, we remove from the current sketch the colors already returned by the previous groups. Thus, the same color is never counted twice.

More precisely, consider a vertex $v\in B_i$. Let $c_v\in\mathbb{R}^{\Dout+1}$ be its color vector. Initially, every coordinate is set to $1$. Whenever the stream contains an edge $(v,u)$ such that $u$ has already been colored with color $k$, we subtract $1$ from coordinate $k$. Thus, a positive coordinate corresponds to an available color, a zero coordinate corresponds
to a color used exactly once by an already colored out-neighbor, and a
negative coordinate corresponds to a color used by at least two already colored out-neighbors.

For every group of $c\log n$ sketches, we additionally remove from the
current sketches all colors returned by the previous groups. This ensures that previously discovered available colors are not sampled again. We also remove the contribution of these colors from the sketches so that the number of negative coordinates does not increase when passing from one group to the next.

We now prove that, with high probability, $L(v)$ contains at least
$\deg_{H_i}(v)+1$ available colors. In fact, we prove the following stronger statement by induction. For every
\[
1\leq j\leq \deg_{H_i}(v)+1,
\]
the first $j c\log n$ sketches return at least $j$ distinct available
colors with probability at least $1-\frac{j}{n^c}.$

For the base case $j=1$, consider the first $c\log n$ sketches. The
probability that an $\ell_0$-sample returns an available color is greater than $1/2$. Indeed, let $x_0,y_0,z_0$ denote, respectively, the number of negative, zero, and positive coordinates of the original color vector.
Here, $x_0$ is the number of colors used by at least two already colored out-neighbors of $v$, $y_0$ is the number of colors used by exactly one such out-neighbor, and $z_0$ is the number of colors available before considering the vertices in $H_i$.

Since every color counted by $x_0$ is used at least twice, we have
\[
2x_0+y_0+\deg_{H_i}(v)
\leq \deg^+(v)
<\Delta_{\mathrm{out}}+1.
\]
Moreover,
\[
x_0+y_0+z_0=\Delta_{\mathrm{out}}+1.
\]
Therefore,
\[
z_0>x_0+\deg_{H_i}(v)\geq x_0,
\]
and hence an $\ell_0$-sample returns an available color with probability
\[
\frac{z_0}{x_0+z_0}>\frac12.
\]
Consequently, by taking $c\log n$ independent samples, the probability
that none of them returns an available color is at most $1/n^c$, for a
sufficiently large constant $c$. This proves the base case.

Now assume that the statement holds for $j-1$, where
\[
2\leq j\leq \deg_{H_i}(v)+1.
\]
Condition on the event that the first $(j-1)c\log n$ sketches return at least $j-1$ distinct available colors. If they return at least $j$ colors, there is nothing to prove. Otherwise, suppose that they return exactly $j-1$ colors.

Let $x_j,y_j,z_j$ denote the numbers of negative, zero, and positive
coordinates seen by the sketches in the $j$-th group. Because the colors returned by the previous groups are removed from the current group, and because their contributions are removed as well, we have
\[
x_j=x_0,\qquad
y_j=y_0+j-1,\qquad
z_j=z_0-j+1.
\]
The important point is that the number of negative coordinates remains
$x_0$; only the previously discovered available colors are removed.
Since $z_0>x_0+\deg_{H_i}(v)$ and $j-1\leq \deg_{H_i}(v),$
we obtain
\[
z_j
=z_0-j+1
>x_0+\deg_{H_i}(v)-j+1
\geq x_0.
\]
Thus, conditioned on the previous $j-1$ groups having found exactly
$j-1$ available colors, every sketch in the $j$-th group returns a new
available color with probability
\[
\frac{z_j}{x_j+z_j}>\frac12.
\]
For a sufficiently large constant $c$, the
probability that the $c\log n$ sketches in this group fail to produce a new available color is at most $1/n^c$.

Now, let $P_j$ denote the set of all outputs produced by the first $j \cdot c \log n$ sketches of the algorithm. Thus, the event that the first $j \cdot c \log n$ sketches output at least $j$ colors is equivalent to $|P_j| \ge j$.

We now bound the probability that the first $j \cdot c \log n$ sketches output at least $j$ colors. By the induction hypothesis $\Pr[|P_{j-1}| \ge j-1] \ge 1-\frac{j-1}{n^c}$. Using the law of total probability and conditioning on the outcome of the first $(j-1) \cdot c \log n$ sketches, we obtain
\[
\begin{aligned}
\Pr[|P_j| \ge j]
&\ge
\Pr[|P_j| \ge j : |P_{j-1}| \ge j-1]
\cdot
\Pr[|P_{j-1}| \ge j-1] \\
&=
\Pr[|P_j| \ge j : |P_{j-1}| = j-1]
\cdot
\Pr[|P_{j-1}| = j-1] \\
& 
+\Pr[|P_j| \ge j : |P_{j-1}| \ge j]
\cdot
\Pr[|P_{j-1}| \ge j] \\
&\ge
(1-\frac{1}{n^c}) \cdot (\Pr[|P_{j-1}|=j-1] + \Pr[|P_{j-1}| \ge j]) \\
&\ge
\left(1-\frac{1}{n^c}\right)
\left(1-\frac{j-1}{n^c}\right)
\ge
1-\frac{j}{n^c}.
\end{aligned}
\]

This completes the induction.

The argument holds for every $j\leq \deg_{H_i}(v)+1.$
Since every block contains at most $\sqrt n$ vertices,
$\deg_{H_i}(v)+1\leq \sqrt n.$
Hence, using $c\log n\sqrt n$ sketches for every vertex is sufficient to obtain at least $\deg_{H_i}(v)+1$ distinct available colors for every vertex in $B_i$, with high probability. We also spent another pass to validate the output colors for $v$ and store $\deg_{H_i}(v)+1$ of them in $L(v)$

We now color the vertices of $H_i$ greedily in reverse topological order.
For a vertex $v\in B_i$, all out-neighbors of $v$ that lie outside $B_i$ have already been colored. By construction, $L(v)$ contains at least $\deg_{H_i}(v)+1$ available colors. Since $v$ has only
$\deg_{H_i}(v)$ out-neighbors inside $H_i$, at least one color in $L(v)$ is not used by an out-neighbor of $v$ in $H_i$. Therefore, the greedy procedure produces a proper coloring of $H_i$, while preserving the proper coloring of all previously processed blocks.

It remains to bound the space and number of passes. For every vertex we maintain $O(\sqrt n\log n)$ $\ell_0$-sampling sketches, each requiring polylogarithmic space. Thus, over all $\sqrt n$ vertices of a fixed block, the total space used
by the sketches is $\tilde O(n).$

The edges of the current induced subgraph $H_i$ contain at most
$O(n)$ edges and can also be stored in $\tilde O(n)$ space. Hence the algorithm uses semi-streaming space.

Finally, each block requires two passes over the stream: one pass to
construct the sampling sketches, store the output colors and the edges of $H_i$, and a second pass to validate the sampled colors and find the actual available colors. Since there are $\sqrt n$ blocks, the total number of passes is $O(\sqrt n).$
By choosing $c$ sufficiently large and applying a union bound, all lists are constructed correctly with high probability. Therefore, the algorithm outputs a proper $(\Dout+1)$-coloring in $O(\sqrt n)$ passes using semi-streaming space.
\end{proof}

\dagprobsthm*
\section{Complexity Classes Based on Semi-Streaming Passes and Consequences}\label{sec:complexity}

Let $\Pi$ and $\Pi'$ be graph streaming problems. We say that $\Pi'$ is $\Pi$-Hard if, given a $p$-pass semi-streaming algorithm for $\Pi'$, we can get an $O(p)$-pass algorithm for $\Pi$.   

\begin{conjecture}\label{conj:reachdag}
Any randomized semi-streaming algorithm for \reachdag requires $\Omega(\sqrt{n})$ passes.
\end{conjecture}

\begin{proposition}
    All problems listed in \Cref{sec:probs} and their DAG versions (if non-trivial) are \reachdag-Hard.
\end{proposition}

\begin{proof}
    The statement is immediate for \ssrdag since an algorithm for it also solves \reachdag. It then follows for DAG versions of \ssr, \msr, \ssspuw (and hence, \ssspw), \sptreeuw (and hence, \sptreew), and \dfs (and hence, \lfdfs), and extend to their general versions. For \acyc and \scc, the DAG versions are trivial, so consider their general version. There are easy reductions from \reachdag to these problems: add the edge $(t,s)$ to the input DAG for \reachdag and run the algorithm for \acyc or \scc. For the former, the graph has a cycle iff $s\leadsto t$. For the latter, $s$ and $t$ will be in the same SCC iff $s\leadsto t$. It then trivially follows for \scctopo. It also follows for \topo since \cite{ChakrabartiGMV20} showed that \topo is harder than \acyc modulo one more pass. For \sconn, \cite{ChenLT25} showed a simple reduction from \reach (which is obviously harder than \reachdag): store edges incident on $s$ and $t$ from the stream (takes $\tO(n)$ space) and after the stream, add edges from $t$ to all other nodes and edges to $s$ from all other nodes. Then the resultant graph is strongly connected iff $s\leadsto t$. For unweighted \longpathsize, we show a reduction, which them implies the hardness for the more general versions of the problem. 


Add $2n-2$ new vertices $s_1, \dots, s_{n-1}$ and $t_1, \dots, t_{n-1}$. Add edges $s_1\to s_2\to\ldots \to s_{n-1}\to s$ and edges $t\to t_1\to t_2\to\ldots \to t_{n-1}$. The new graph contains a path of length at least $2n-1$ if and only if $t$ is reachable from $s$ in $G$. If $t$ is reachable from $s$, then there is a path of length at least $(n-1)+1+(n-1)=2n-1$ from $s_1$ to $t_{n-1}$. Any path in the original graph that is not an $s,t$-path can connect at most one of the two new chains added, and hence, if $t$ is not reachable from $s$, then length of any path in the new graph is at most $(n-1)+(n-1)=2n-2$.

This completes the proof of the proposition.
\end{proof}

Since all our algorithms for the DAG problems use $n^{1/2+o(1)}$ passes, the following proposition follows.

\begin{proposition}
    If \Cref{conj:reachdag} is true, then all our algorithms are optimal up to a factor of $n^{o(1)}$ in passes. This means the semi-streaming pass complexity of all problems listed in \Cref{sec:probs} is $\Theta^
    \star(\sqrt{n})$.  
\end{proposition}

Since \topo is \reachdag-Hard, the following conjecture is weaker than \Cref{conj:reachdag}.

\begin{conjecture}\label{conj:topo}
Any randomized semi-streaming algorithm for \topo requires $\Omega(\sqrt{n})$ passes.
\end{conjecture}

\begin{proposition}
   The problems \scctopo, \ssspwdag, \sptreewdag, \dfsfindag, and \lfdfsdag are \topo-Hard.
\end{proposition}

\begin{proof}
The claim about \scctopo is trivial since it precisely solves \topo on a DAG. For reduction to \ssspwdag, add a source $s$ and edges of weight $0$ from $s$ to all nodes in the original graph, and assign every original edge a weight of $-1$. Then $\text{dist}(s,v)=-\ell(v)$ where $\ell(v)$ is the maximum number of edges in a path ending at $v$. Thus, for each edge $u\to v$ in the original graph, $\text{dist}(s,u)>\text{dist}(s,v)$. Hence, we can order the nodes by decreasing distance from $s$ and obtain a topological order. The argument of \sptreewdag is the same since it also gives the distance from $s$ to every node. Finally, \dfsfindag can be used to obtain a topological ordering by sorting by decreasing order of finish time (the classical textbook algorithm), and \lfdfsdag gives the finish times if we run DFS on the returned \lfdfs.
\end{proof}

\begin{proposition}
    If \Cref{conj:topo} is true, then our algorithms for \ssspwdag, \sptreewdag, \dfsfindag, and \lfdfsdag are optimal up to a factor of $O(1)$ in passes. Hence, all these problems are \topo-complete.   
\end{proposition}

\begin{proof}
Let $p_{\topo}$ be the optimal number of semi-streaming passes required to solve \topo. By \Cref{thm:dagprobs}, our algorithms solve these problems using $O(p_{\topo}+\sqrt{n})$ passes. If \Cref{conj:topo} is true, then $\sqrt{n}=O(p_{\topo})$, which means we solve these problems in $O(p_{\topo})$ passes. Since the problems are \topo-Hard, they need $\Omega(p_{\topo})$ passes by definition and it also follows that they are \topo-complete. Thus, the proposition follows.     
\end{proof}

\begin{conjecture}\label{conj:ssr}
Any randomized semi-streaming algorithm for \ssr requires $\Omega(\sqrt{n})$ passes.
\end{conjecture}

\begin{proposition}
   The problems \ssspuwdag, \ssspposwdag, \sptreeuwdag, \sptreeposwdag, and \lfdfsdag are \ssr-Hard.
\end{proposition}

\begin{proof}
    All but \lfdfsdag not only tells whether each vertex is reachable from the source node but also gives the distances ($\infty$ if unreachable). For \lfdfsdag, if we keep the source first in the lexicographic ordering, then it will be first root and the DFS forest returned will contain the DFS tree rooted at $s$, which in turn contains all nodes reachable from $s$.
\end{proof}

\begin{proposition}
    If \Cref{conj:ssr} is true, then our algorithms for \ssspuwdag, \ssspposwdag, \sptreeuwdag, \sptreeposwdag, and \lfdfsdag are optimal up to a factor of $O(\log n)$ in passes.   
\end{proposition}

\begin{proof}
    Let $p_{\topo}$ be defined as above. By \Cref{thm:dagprobs}, our algorithms solve these problems using $O(p_{\ssr}\cdot \log n+\sqrt{n})$ passes. If \Cref{conj:ssr} is true, then $\sqrt{n}=O(p_{\ssr})$, which means we solve these problems in $O(p_{\ssr}\cdot\log n)$ passes. Since the problems are \ssr-Hard, they need $\Omega(p_{\ssr})$ passes by definition. Thus, the proposition follows. 
\end{proof}

\section{AI Disclosure Statement}

We used ChatGPT 5.6 by OpenAI to assist with brainstorming on the problems. The tool materially affected \Cref{sec:dfs} and \Cref{sec:lbs}. All proofs are written by the author(s) and they take full responsibility of the material in the article. The author(s) verified the correctness and originality of all content including references.



\bibliography{refs}

@Article{Ablayev96,
  author =       {Farid Ablayev},
  title =        {Lower Bounds for One-way Probabilistic Communication Complexity and Their Application to Space Complexity},
  journal =      {Theoretical Computer Science},
  year =         1996,
  volume =       175,
  number =       2,
  pages =        {139--159}
}

@inproceedings{BeraCG19,
  author    = {Suman K. Bera and
               Amit Chakrabarti and
               Prantar Ghosh},
  title     = {Graph Coloring via Degeneracy in Streaming and Other Space-Conscious
               Models},
  booktitle = {47th International Colloquium on Automata, Languages, and Programming,
               {ICALP} 2020, July 8-11, 2020, Saarbr{\"{u}}cken, Germany (Virtual
               Conference)},
  pages     = {11:1--11:21},
  year      = {2020}
  }

@inproceedings{BCKWY14,
  author    = {Joshua Brody and Amit Chakrabarti and Ranganath Kondapally and David P. Woodruff and Grigory Yaroslavtsev},
  title     = {Certifying Equality With Limited Interaction},
  booktitle = {Approximation, Randomization, and Combinatorial Optimization. Algorithms and Techniques (APPROX/RANDOM 2014)},
  series    = {Leibniz International Proceedings in Informatics (LIPIcs)},
  volume    = {28},
  pages     = {545--581},
  year      = {2014},
  publisher = {Schloss Dagstuhl -- Leibniz-Zentrum f{\"u}r Informatik},
  doi       = {10.4230/LIPIcs.APPROX-RANDOM.2014.545}
}

@inproceedings{JLS19,
  author    = {Arun Jambulapati and Yang P. Liu and Aaron Sidford},
  title     = {Parallel Reachability in Almost Linear Work and Square Root Depth},
  booktitle = {Proceedings of the 60th IEEE Annual Symposium on Foundations of Computer Science (FOCS 2019)},
  editor    = {David Zuckerman},
  pages     = {1664--1686},
  year      = {2019},
  publisher = {IEEE Computer Society},
  doi       = {10.1109/FOCS.2019.00098}
}

@thesis{Janthong14,
  author      = {Abhabongse Janthong},
  title       = {Streaming Algorithm for Determining a Topological Ordering of a Digraph},
  type        = {Undergraduate Thesis},
  institution = {Brown University},
  year        = {2014},
  month       = apr,
  note        = {Advisor: Paul Valiant}
}

@article{AzarmehrBCGLR26,
  author  = {Amir Azarmehr and Soheil Behnezhad and Lily Chung and Alma Ghafari and Jane Lange and Ronitt Rubinfeld},
  title   = {Streaming Algorithms for Monotonicity Testing},
  journal = {arXiv preprint arXiv:2608.07073},
  year    = {2026},
  doi     = {10.48550/arXiv.2608.07073},
  url     = {https://arxiv.org/abs/2608.07073}
}

@inproceedings{GhoshKMM26,
  author    = {Prantar Ghosh and Sahil Kuchlous and Shravan Mehra and Sagnik Mukhopadhyay},
  title     = {The Power of the Score Sequence of a Tournament},
  booktitle = {34th Annual European Symposium on Algorithms (ESA 2026)},
  series    = {Leibniz International Proceedings in Informatics (LIPIcs)},
  volume    = {388},
  pages     = {156:1--156:19},
  year      = {2026},
  publisher = {Schloss Dagstuhl -- Leibniz-Zentrum f{\"u}r Informatik},
  doi       = {10.4230/LIPIcs.ESA.2026.156}
}

@inproceedings{KelleyW59,
  author    = {Kelley, James E. and Morgan R. Walker},
  title     = {Critical-Path Planning and Scheduling},
  booktitle = {Papers Presented at the December 1--3, 1959, Eastern Joint IRE-AIEE-ACM Computer Conference},
  pages     = {160--173},
  year      = {1959},
  publisher = {Association for Computing Machinery},
  doi       = {10.1145/1460299.1460318}
}

@inproceedings{AKM22,
  author    = {Sepehr Assadi and Pankaj Kumar and Parth Mittal},
  title     = {Brooks' Theorem in Graph Streams: A Single-Pass Semi-Streaming Algorithm for {$\Delta$}-Coloring},
  booktitle = {Proceedings of the 54th Annual ACM SIGACT Symposium on Theory of Computing (STOC 2022)},
  pages     = {234--247},
  year      = {2022},
  publisher = {Association for Computing Machinery},
  doi       = {10.1145/3519935.3520005}
}

@inproceedings{KSS16,
  author    = {Lasse Kliemann and Christian Schielke and Anand Srivastav},
  title     = {A Streaming Algorithm for the Undirected Longest Path Problem},
  booktitle = {24th Annual European Symposium on Algorithms (ESA 2016)},
  series    = {Leibniz International Proceedings in Informatics (LIPIcs)},
  volume    = {57},
  pages     = {56:1--56:17},
  year      = {2016},
  publisher = {Schloss Dagstuhl -- Leibniz-Zentrum f{\"u}r Informatik},
  doi       = {10.4230/LIPIcs.ESA.2016.56}
}

@misc{ChenCLT26,
      title={Independence-Number Parameterized Space Complexity for Directed Connectivity Certificate}, 
      author={Ho-Lin Chen and Tsun Ming Cheung and Peng-Ting Lin and Meng-Tsung Tsai},
      year={2026},
      eprint={2602.12668},
      archivePrefix={arXiv},
      primaryClass={cs.DS},
      url={https://arxiv.org/abs/2602.12668}, 
}

@inproceedings{AssadiJJST22,
  author    = {Sepehr Assadi and
               Arun Jambulapati and
               Yujia Jin and
               Aaron Sidford and
               Kevin Tian},
  title     = {Semi-Streaming Bipartite Matching in Fewer Passes and Optimal Space},
  booktitle = {{SODA}},
  pages     = {627--669},
  publisher = {{SIAM}},
  year      = {2022}
}

@inproceedings{AssadiR20,
  author    = {Sepehr Assadi and
               Ran Raz},
  title     = {Near-Quadratic Lower Bounds for Two-Pass Graph Streaming Algorithms},
  booktitle = {{FOCS}},
  pages     = {342--353},
  publisher = {{IEEE}},
  year      = {2020}
}

@article{FeigenbaumKMSZ05,
  author    = {Joan Feigenbaum and
               Sampath Kannan and
               Andrew McGregor and
               Siddharth Suri and
               Jian Zhang},
  title     = {On graph problems in a semi-streaming model},
  journal   = {Theor. Comput. Sci.},
  volume    = {348},
  number    = {2-3},
  pages     = {207--216},
  year      = {2005}
}

@inproceedings{ChangFHT20,
  author    = {Yi{-}Jun Chang and
               Martin Farach{-}Colton and
               Tsan{-}sheng Hsu and
               Meng{-}Tsung Tsai},
  title     = {Streaming Complexity of Spanning Tree Computation},
  booktitle = {37th International Symposium on Theoretical Aspects of Computer Science,
               {STACS} 2020, March 10-13, 2020, Montpellier, France},
  pages     = {34:1--34:19},
  year      = {2020}
  }

@inproceedings{KhanM19,
  author    = {Shahbaz Khan and
               Shashank K. Mehta},
  title     = {Depth First Search in the Semi-streaming Model},
  booktitle = {36th International Symposium on Theoretical Aspects of Computer Science,
               {STACS} 2019, March 13-16, 2019, Berlin, Germany},
  pages     = {42:1--42:16},
  year      = {2019}
  }

@inproceedings{AssadiCK19a,
  author    = {Sepehr Assadi and
               Yu Chen and
               Sanjeev Khanna},
  title     = {Sublinear Algorithms for ({\(\Delta\)} + 1) Vertex Coloring},
  booktitle = {Proceedings of the Thirtieth Annual {ACM-SIAM} Symposium on Discrete
               Algorithms, {SODA} 2019, San Diego, California, USA, January 6-9,
               2019},
  pages     = {767--786},
  year      = {2019}
  }

@inproceedings{GuruswamiO13,
  author    = {Venkatesan Guruswami and
               Krzysztof Onak},
  title     = {Superlinear Lower Bounds for Multipass Graph Processing},
  booktitle = {Proceedings of the 28th Conference on Computational Complexity, {CCC}
               2013, K.lo Alto, California, USA, 5-7 June, 2013},
  pages     = {287--298},
  year      = {2013}
  }

@inproceedings{ChakrabartiGMV20,
  author    = {Amit Chakrabarti and
               Prantar Ghosh and
               Andrew McGregor and
               Sofya Vorotnikova},
  title     = {Vertex Ordering Problems in Directed Graph Streams},
  booktitle = {Proceedings of the 2020 {ACM-SIAM} Symposium on Discrete Algorithms,
               {SODA} 2020, Salt Lake City, UT, USA, January 5-8, 2020},
  pages     = {1786--1802},
  year      = {2020}
  }

@inproceedings{JowhariST2011,
  title={Tight bounds for Lp samplers, finding duplicates in streams, and related problems},
  author={Jowhari, Hossein and Sa{\u{g}}lam, Mert and Tardos, G{\'a}bor},
  booktitle={Proceedings of the thirtieth ACM SIGMOD-SIGACT-SIGART symposium on Principles of database systems},
  pages={49--58},
  year={2011},
  organization={ACM}
}

@inproceedings{ChakrabartiSWY01,
  author    = {Amit Chakrabarti and
               Yaoyun Shi and
               Anthony Wirth and
               Andrew Chi{-}Chih Yao},
  title     = {Informational Complexity and the Direct Sum Problem for Simultaneous
               Message Complexity},
  booktitle = {42nd Annual Symposium on Foundations of Computer Science, {FOCS} 2001,
               14-17 October 2001},
  pages     = {270--278},
  year      = {2001}
  }

@inproceedings{Yao79,
  author    = {Andrew Chi{-}Chih Yao},
  title     = {Some Complexity Questions Related to Distributive Computing (Preliminary
               Report)},
  booktitle = {Proceedings of the 11h Annual {ACM} Symposium on Theory of Computing,
               April 30 - May 2, 1979, Atlanta, Georgia, {USA}},
  pages     = {209--213},
  year      = {1979}
  }

@book{DubhashiP09,
  author    = {Devdatt P. Dubhashi and
               Alessandro Panconesi},
  title     = {Concentration of Measure for the Analysis of Randomized Algorithms},
  publisher = {Cambridge University Press},
  year      = {2009}
  }

@article{Elkin11,
  author    = {Michael Elkin},
  title     = {Streaming and fully dynamic centralized algorithms for constructing
               and maintaining sparse spanners},
  journal   = {{ACM} Trans. Algorithms},
  volume    = {7},
  number    = {2},
  pages     = {20:1--20:17},
  year      = {2011}
  }

@inproceedings{ChenLT25,
  author    = {Ho-Lin Chen and Peng-Ting Lin and Meng-Tsung Tsai},
  title     = {Parameterized Streaming Algorithms for Topological Sorting},
  booktitle = {19th International Symposium on Algorithms and Data Structures
               (WADS 2025)},
  series    = {Leibniz International Proceedings in Informatics (LIPIcs)},
  volume    = {349},
  pages     = {18:1--18:20},
  year      = {2025},
  publisher = {Schloss Dagstuhl -- Leibniz-Zentrum f{\"u}r Informatik},
  doi       = {10.4230/LIPIcs.WADS.2025.18}
}

@inproceedings{ChenKPSSY21,
  author    = {Lijie Chen and Gillat Kol and Dmitry Paramonov and
               Raghuvansh R. Saxena and Zhao Song and Huacheng Yu},
  title     = {Almost Optimal Super-Constant-Pass Streaming Lower Bounds
               for Reachability},
  booktitle = {Proceedings of the 53rd Annual ACM SIGACT Symposium on
               Theory of Computing (STOC 2021)},
  pages     = {570--583},
  year      = {2021},
  publisher = {ACM},
  doi       = {10.1145/3406325.3451038}
}

@inproceedings{LauraS11,
  author    = {Luigi Laura and Federico Santaroni},
  title     = {Computing Strongly Connected Components in the Streaming Model},
  booktitle = {Theory and Practice of Algorithms in (Computer) Systems
               (TAPAS 2011)},
  series    = {Lecture Notes in Computer Science},
  volume    = {6595},
  pages     = {193--205},
  year      = {2011},
  publisher = {Springer},
  doi       = {10.1007/978-3-642-19754-3_20}
}

@inproceedings{BawejaJW22,
  author    = {Anubhav Baweja and Justin Jia and David P. Woodruff},
  title     = {An Efficient Semi-Streaming {PTAS} for Tournament Feedback
               Arc Set with Few Passes},
  booktitle = {13th Innovations in Theoretical Computer Science Conference
               (ITCS 2022)},
  series    = {Leibniz International Proceedings in Informatics (LIPIcs)},
  volume    = {215},
  pages     = {16:1--16:23},
  year      = {2022},
  publisher = {Schloss Dagstuhl -- Leibniz-Zentrum f{\"u}r Informatik},
  doi       = {10.4230/LIPIcs.ITCS.2022.16}
}

@inproceedings{GhoshK24,
  author    = {Prantar Ghosh and Sahil Kuchlous},
  title     = {New Algorithms and Lower Bounds for Streaming Tournaments},
  booktitle = {32nd Annual European Symposium on Algorithms (ESA 2024)},
  series    = {Leibniz International Proceedings in Informatics (LIPIcs)},
  volume    = {308},
  pages     = {60:1--60:19},
  year      = {2024},
  publisher = {Schloss Dagstuhl -- Leibniz-Zentrum f{\"u}r Informatik},
  doi       = {10.4230/LIPIcs.ESA.2024.60}
}

@inproceedings{AssadiHS26,
  author    = {Sepehr Assadi and Gary Hoppenworth and Janani Sundaresan},
  title     = {Better Bounds for Semi-Streaming Single-Source Shortest Paths},
  booktitle = {Proceedings of the 2026 Annual ACM-SIAM Symposium on
               Discrete Algorithms (SODA 2026)},
  pages     = {5093--5116},
  year      = {2026},
  publisher = {SIAM},
  doi       = {10.1137/1.9781611978971.184}
}

@inproceedings{KonradT25,
  author    = {Christian Konrad and Chhaya Trehan},
  title     = {Constructing Long Paths in Graph Streams},
  booktitle = {33rd Annual European Symposium on Algorithms (ESA 2025)},
  series    = {Leibniz International Proceedings in Informatics (LIPIcs)},
  volume    = {351},
  pages     = {22:1--22:19},
  year      = {2025},
  publisher = {Schloss Dagstuhl -- Leibniz-Zentrum f{\"u}r Informatik},
  doi       = {10.4230/LIPIcs.ESA.2025.22}
}

@inproceedings{AssadiGLMM24,
  author    = {Sepehr Assadi and Prantar Ghosh and Bruno Loff and
               Parth Mittal and Sagnik Mukhopadhyay},
  title     = {Polynomial Pass Semi-Streaming Lower Bounds for
               {$K$}-Cores and Degeneracy},
  booktitle = {39th Computational Complexity Conference (CCC 2024)},
  series    = {Leibniz International Proceedings in Informatics (LIPIcs)},
  volume    = {300},
  pages     = {7:1--7:16},
  year      = {2024},
  publisher = {Schloss Dagstuhl -- Leibniz-Zentrum f{\"u}r Informatik},
  doi       = {10.4230/LIPIcs.CCC.2024.7}
}

@book{CLRS,
  author    = {Thomas H. Cormen and Charles E. Leiserson and Ronald L. Rivest and Clifford Stein},
  title     = {Introduction to Algorithms},
  edition   = {4},
  year      = {2022},
  publisher = {MIT Press},
  address   = {Cambridge, MA}
}

\appendix

\section{Lower Bounds for $(\Dout+1)$-coloring}\label{sec:lbs}

In this section, we study lower bounds for semi-streaming algorithms for proper $(\Dout+1)$-coloring. We first show an $\Omega(n^2)$ space lower bound for one-pass algorithms. In fact, the reduction continues to work even when the topological ordering of the graph and the value of $\Dout$ are given to the algorithm before the stream starts.

We then consider the multi-pass setting. In this case, we allow the stream to contain multiple copies of the same edge and show an
$\Omega(n^2/p)$ space lower bound for a $p$-pass algorithm.

Before proving the results, we discuss communication complexity preliminaries that we use for the proofs.

\paragraph{Communication Complexity with Applications to Streaming Lower Bounds.} In the two-player communication model of Yao \cite{yao79}, the players Alice and Bob receive $\bx\in \{0,1\}^N$ and $\by\in \{0,1\}^N$ respectively, and the goal is to communicate as few bits as possible to output $f(\bx,\by)$ for a given function $f$. We consider the {\em randomized setting} where the players may use private random coins to generate the messages. The \emph{communication cost} of a randomized protocol for $f$ is the maximum number of bits exchanged by the players, over all possible inputs $(\bx,\by)$ and all possible random strings. The \emph{randomized communication complexity} of a function $f$, denoted by $R(f)$, is the minimum communication cost over all randomized protocols (allowing unlimited rounds of interaction) that, for all possible input pairs $\bx,\by$, output $f(\bx,\by)$ correctly with probability at least $2/3$. 

A restrictive variant of the model is the \emph{one-way communication model} where Alice sends one message to Bob, following which he must output $f(\bx,\by)$. The \emph{one-way randomized communication complexity} of $f$, denoted by $R^{\to}(f)$, is defined as the minimum communication cost over all randomized one-way protocols.

    To prove our lower bounds, we reduce from the {\em Index} (\idx) and (a generalized version of) the {\em Unique Set Intersection} (\usetint) problems. We formally define them below and mention the results on their communication complexity that we use.  
    \begin{description}
        \item $\idx_N$: Alice gets $\bx \in \{0, 1\}^N$ and Bob gets an index $i \in [N]$. The goal is to find $\bx_i$, the $i$th bit of $\bx$.
    \end{description}

         \begin{fact}[\cite{Ablayev96}] \label{fact:idx-lb}
        The one-way randomized communication complexity $R^{\to}(\idx_N) = \Omega(N)$ .
    \end{fact}

\begin{description}
        \item $\usetint_N$: Alice gets $\bx \in \{0, 1\}^N$ and Bob gets $\by \in \{0, 1\}^N$, where $\bx$ and $\by$ are promised to intersect uniquely as sets, i.e., there is a unique $i \in [N]$ such that $\bx_i = \by_i = 1$. The goal is to output $i$.
\end{description}

\noindent
        We actually use the communication complexity of {\em $k$-fold Unique Set Intersection} problem $k$-\usetint, where Alice and Bob are given $k$ independent instances of $\usetint_N$ for some $k$ and $N$. The ``direct sum'' theorem says that its communication complexity is $\Omega(k)$ times that of $\usetint_N$.

 \begin{fact}[Direct sum theorem \cite{chakrabartiswy01}]\label{fact:dirsum}
    Let $f$ be a function such that Alice and Bob need to compute {\em each} of $f(x_1,y_1),\ldots, f(x_k,y_k)$ with probability at least $2/3$. Then the required communication is $\Omega(k)\cdot R(f)$.    
\end{fact}

Since $R(\usetint_N)=\Omega(N)$ \cite{BCKWY14}, the following fact follows from \Cref{fact:dirsum}.

\begin{fact}\label{fact:setint-lb}
        The randomized communication complexity $R(k$-$\usetint_N)=\Omega(kN)$.
\end{fact}

\subsection{One-Pass Space Lower Bound.}

\begin{theorem}
Let $\mathcal{A}$ be a randomized one-pass insertion-only streaming
algorithm that, given an $n$-vertex DAG $G$, outputs with probability at least $2/3$ a proper $(\Dout+1)$-coloring. Then $\mathcal{A}$ requires $\Omega(n^2)$ space.
\end{theorem}

We prove the theorem by a reduction from the one-way communication problem \textsc{INDEX}. In an instance of \textsc{INDEX}, Alice has a binary matrix $X\in\{0,1\}^{p\times p}$ and Bob has a pair $(y,z)\in[p]\times[p]$.
Bob's goal is to find the entry $X_{y,z}$. It is known that any randomized one-way protocol for \textsc{INDEX} with constant success probability requires $\Omega(p^2)$ bits of communication.

We first describe the reduction.

\begin{lemma}
Suppose there is a randomized one-pass insertion-only streaming algorithm $\mathcal{A}$ that computes a proper $(\Dout+1)$-coloring of every $n$-vertex DAG using $s$ bits of space. Then there is a one-way communication protocol that solves a $p\times p$ instance of
\textsc{INDEX} using $s$ bits of communication, where $n=4p+1$.
\end{lemma}

\begin{proof}
Suppose Alice is given a binary matrix
$X=(x_{i,j})_{i,j\in[p]}$. She constructs a graph with $4p+1$ vertices, partitioned into the following four sets:
$$
V=\{v_1,\ldots,v_p\},
U=\{u_1,\ldots,u_p\},
W=\{w_1,\ldots,w_p\},
$$
and
$$
S=\{s_1,\ldots,s_{p+1}\}.
$$

For every $i,j\in[p]$, Alice adds exactly one of the two edges
$(v_i,u_j)$ and $(v_i,w_j)$ according to the value of $x_{i,j}$:
$$
(v_i,u_j)\in E \quad\text{if }x_{i,j}=1,
$$
and
$$
(v_i,w_j)\in E \quad\text{if }x_{i,j}=0.
$$
She also adds a directed clique on $S$, where the edges are oriented
according to the order
$$
s_1,s_2,\ldots,s_{p+1}.
$$
That is, $(s_i,s_j)$ is added for every $i<j$.

Alice now runs $\mathcal{A}$ on the edges she constructed. After processing her part of the stream, she sends the current memory state of $\mathcal{A}$ to Bob.

Suppose Bob's input is $(y,z)$. He defines a mapping
$f:U\cup W\rightarrow S$ by
$$
f(u_j)=f(w_j)=s_j\qquad\text{for }j\neq z,
$$
and
$$
f(u_z)=s_z,\qquad f(w_z)=s_{p+1}.
$$
Bob then continues the execution of $\mathcal{A}$. For every
$a\in U\cup W$, he adds an edge from $a$ to every vertex in $S$ except
$f(a)$. Thus, he adds
$$
(a,s)\qquad\text{for every }s\in S\setminus\{f(a)\}.
$$

The directed clique on $S$ contains $p+1$ vertices. Hence, in every proper $(\Dout+1)$-coloring, all vertices in $S$ must receive different colors. Also, the maximum out-degree of the resulting graph is $p$, and therefore $\Dout+1=p+1$.

Consider first $j\neq z$. Both $u_j$ and $w_j$ have an edge to every
vertex of $S$ except $s_j$. Since all vertices in $S$ have different
colors and there are exactly $p+1$ available colors, the only possible
color for both $u_j$ and $w_j$ is the color of $s_j$. Thus,
$$
c(u_j)=c(w_j)=c(s_j)\qquad\text{for every }j\neq z.
$$

For the special index $z$, the same argument gives
$c(u_z)=c(s_z)$
and $c(w_z)=c(s_{p+1}).$

We now consider $v_y$. For every $j\neq z$, Alice added either
$(v_y,u_j)$ or $(v_y,w_j)$. In both cases, the corresponding vertex has color $c(s_j)$. Therefore, $v_y$ cannot use any of the colors
$c(s_j)$ for $j\neq z$. The only two possible colors for $v_y$ are
$c(s_z)$ and $c(s_{p+1})$.

If $x_{y,z}=1$, then $(v_y,u_z)$ is present. Since
$c(u_z)=c(s_z)$, $v_y$ cannot use $c(s_z)$, and hence
$c(v_y)=c(s_{p+1}).$ On the other hand, if $x_{y,z}=0$, then $(v_y,w_z)$ is present. Since $c(w_z)=c(s_{p+1})$, we must have
$c(v_y)=c(s_z).$ Therefore, Bob can recover $x_{y,z}$ by comparing the color of $v_y$ with the colors of $s_z$ and $s_{p+1}$:
$$
x_{y,z}=
\begin{cases}
1,&\text{if }c(v_y)=c(s_{p+1}),\\
0,&\text{if }c(v_y)=c(s_z).
\end{cases}
$$

It remains to check that the graph is a DAG. One valid topological ordering is obtained by placing the four sets in the order
$$
V,\;U,\;W,\;S,
$$
and ordering the vertices of $S$ as
$s_1,\ldots,s_{p+1}$. All edges added by Alice go from $V$ to $U\cup W$
or are edges inside $S$. All edges added by Bob go from $U\cup W$ to $S$. Thus, this ordering is a valid topological ordering.

The graph has $n=4p+1=\Theta(p)$ vertices. This proves the lemma.
\end{proof}

\begin{proof}[Proof of Theorem 3.1]
The one-way communication complexity of a $p\times p$ instance of
\textsc{INDEX} is $\Omega(p^2)$. By the lemma, an $s$-space streaming
algorithm would give a one-way protocol using $s$ bits of communication.
Therefore,
$$
s=\Omega(p^2)=\Omega(n^2),
$$
since $n=\Theta(p)$. The theorem follows.
\end{proof}

\begin{remark}
The reduction also shows that the lower bound holds even if the
topological ordering and $\Dout$ are known to the algorithm before the
stream starts. In fact, neither of these pieces of information depends on Alice's matrix $X$. Hence, they can be given to both Alice and Bob in advance without affecting the communication reduction.
\end{remark}

\subsection{Multipass lower bound for repeated edges.}

We next consider the multi-pass setting in which the stream may contain multiple copies of the same edge. We prove a space lower bound by reducing from the $n$-fold Unique Intersection problem.

\begin{theorem}\label{thm:}
Let $\mathcal{A}$ be a randomized $p$-pass streaming algorithm that,
given an $n$-vertex DAG whose stream may contain multiple copies of the same edge, outputs a proper $(\Dout+1)$-coloring with constant success probability. Then $\mathcal{A}$ requires
$\Omega\left(\frac{n^2}{p}\right)$ space.
\end{theorem}

In the $n$-fold Unique Intersection problem, Alice is given $n$ binary
vectors
$$
a_1,\ldots,a_n\in\{0,1\}^n,
$$
and Bob is given another $n$ binary vectors
$$
b_1,\ldots,b_n\in\{0,1\}^n.
$$
It is promised that, for every $i\in[n]$, there is a unique index
$s_i\in[n]$ such that $a_i[s_i]=b_i[s_i]=1.$
The goal is for Bob to recover all indices $s_1,\ldots,s_n$.
We use the following reduction.

\begin{proof}[Proof of ]
Suppose that $\mathcal{A}$ uses $s$ bits of space. We construct a
communication protocol for the $n$-fold Unique Intersection problem.

Let Alice have $a_1,\ldots,a_n$ and Bob have $b_1,\ldots,b_n$, satisfying the unique-intersection promise. The graph has $2n$ vertices, divided into two sets
$$
L=\{l_1,\ldots,l_n\}
\ \text{and}\
R=\{r_1,\ldots,r_n\}.
$$

Alice first constructs the part of the stream determined by her input.
For every $i,j\in[n]$ such that $a_i[j]=0,$ she inserts the edge $(l_i,r_j).$ She then simulates $\mathcal{A}$ on this part of the stream.

Since $\mathcal{A}$ makes multiple passes over the stream, Alice and Bob simulate the passes one by one. At the end of each pass, Alice sends the current memory state of $\mathcal{A}$ to Bob, and Bob continues the same pass on his part of the stream. At the end of Bob's part, the current memory state is sent back to Alice so that she can start the next pass. Thus, each pass requires $O(s)$ bits of communication, and the complete simulation uses $O(ps)$ bits.

During his part of the stream, Bob inserts an edge $(l_i,r_j)$ whenever $b_i[j]=0.$ Notice that an edge can occur in both Alice's and Bob's parts of the stream. This is exactly where we use the fact that the streaming model allows multiple copies of the same edge.

After processing the edges determined by his vectors, Bob adds the edges $(r_i,r_j)$ for every $ i<j.$ Therefore, the vertices in $R$ induce a directed clique.

We now consider the coloring produced by $\mathcal{A}$. By the promise, for every $i\in[n]$, there is a unique index $s_i$ for which
$a_i[s_i]=b_i[s_i]=1.$ Hence, for every $j\neq s_i$, at least one of $a_i[j]$ and $b_i[j]$ is zero. Therefore, the edge $(l_i,r_j)$ appears in the stream for every $j\neq s_i$. On the other hand, $(l_i,r_{s_i})$ never appears.

Consequently,$N^+(l_i)=R\setminus\{r_{s_i}\},$ and hence $d^+(l_i)=n-1.$
The directed clique on $R$ also implies that $\Dout=n-1$ and all vertices of $R$ receive different colors in every proper coloring. Thus, the algorithm uses exactly $n$ available colors.

It follows that $l_i$ is adjacent to $n-1$ vertices having pairwise
different colors. The only color that can be assigned to $l_i$ is
therefore the color of the unique vertex $r_{s_i}$ that is not an
out-neighbor of $l_i$. Hence, $c(l_i)=c(r_{s_i}).$ Since all vertices of $R$ have distinct colors, Bob can recover $s_i$ by comparing the color of $l_i$ with the colors of the vertices in $R$. Doing this for every $i$ allows Bob to recover $s_1,\ldots,s_n.$

We finally check that the graph constructed in the reduction is a DAG.
All edges between $L$ and $R$ are directed from $L$ to $R$, and the edges inside $R$ are directed according to the ordering $r_1,r_2,\ldots,r_n.$
Thus, $l_1,\ldots,l_n,r_1,\ldots,r_n$ is a valid topological ordering.

The reduction gives a communication protocol for the $n$-fold Unique
Intersection problem using $O(ps)$ bits. The communication complexity of this problem is $\Omega(n^2)$. Therefore, $ps=\Omega(n^2),$
which gives $s=\Omega\left(\frac{n^2}{p}\right).$ This proves the theorem.
\end{proof}

\end{document}